\documentclass[onecolumn]{emulateapj}

\usepackage{epsfig}    
\usepackage[figuresright]{rotating}  
\usepackage{amsmath}   
\usepackage{amssymb}   
\usepackage{latexsym}  
\usepackage{dcolumn}
\newcolumntype{.}{D{.}{.}{-1}}

\def\etal{{\em et~al.~}}
\def\vec#1{{\bf #1}}
\def\mvec#1{\boldsymbol{#1}}

\begin{document}

\title{Scattering polarization and Hanle effect in stellar atmospheres with horizontal inhomogeneities}

\shorttitle{Scattering polarization in inhomogeneous atmospheres}

\author{Rafael Manso Sainz\altaffilmark{1, 2} \& Javier Trujillo Bueno\altaffilmark{1, 2, 3}}
\altaffiltext{1}{Instituto de Astrof\'{\i}sica de Canarias, E-38205, La Laguna, 
Tenerife, Spain}
\altaffiltext{2}{Departamento de Astrof\'{\i}sica, Facultad de F\'{\i}sica, Universidad de La Laguna, Tenerife, Spain}
\altaffiltext{3}{Consejo Superior de Investigaciones Cient\'{\i}ficas, Spain}
\email{rsainz@iac.es, jtb@iac.es}


\begin{abstract}

Scattering of light from an anisotropic source produces linear polarization in spectral lines and the continuum. 
In the outer layers of a stellar atmosphere the anisotropy of the radiation field is typically dominated by the
radiation escaping away, but local horizontal fluctuations of the physical conditions may also contribute, 
distorting the illumination and hence, the polarization pattern. Additionally, a magnetic field may perturb and modify the line scattering polarization signals through the Hanle effect. Here, we study such symmetry-breaking effects. We develop a method to solve the transfer of polarized radiation in a scattering atmosphere with {\em weak} horizontal fluctuations of the opacity and source functions. It comprises linearization (small opacity fluctuations are assumed), reduction to a quasi-planeparallel problem through harmonic analysis,
and numerical solution by generalized standard techniques. We apply this method to study scattering polarization in atmospheres
with horizontal fluctuations in the Planck function and opacity. We derive several very general results and constraints from considerations on the symmetries and dimensionality of the problem, and we give explicit solutions of a few illustrative problems of especial interest. For example, we show (a) how the amplitudes of the fractional linear polarization signals change when considering increasingly smaller horizontal atmospheric inhomogeneities, (b) that in the presence of such inhomogeneities even a vertical magnetic field may modify the scattering line polarization, and (c) that forward scattering polarization may be produced without the need of an inclined magnetic field. These results are important to understand the physics of the problem and as benchmarks for multidimensional radiative transfer codes.


\end{abstract}


\section{Introduction}\label{sect1}

The polarization generated by scattering in a spectral line depends sensitively on the global geometry of the scattering process (most notably, the distribution of incident radiation), as well as on the angular momentum of the atomic levels involved in the transition.
If the incident illumination is isotropic, the scattered radiation is unpolarized;
if the illumination is collimated, the polarization of scattered
light may be significant, reaching complete polarization at 90$^\circ$ of the
incident beam in a $0\rightarrow 1$ transition (total angular momentum $J=0, 1$ for
the lower and upper levels of the transition, respectively ).
In the outer layers of a stellar atmosphere the radiation field is anisotropic mainly 
because light is escaping through the surface.
When quantified (Sect.~\ref{sect2} below), the degree of anisotropy 
in a plane-parallel, semi-infinite atmosphere is typically of a few percent, which yields
signals with polarization degrees on that order of magnitude or lower 
(Stenflo et al. 1983a, b; Stenflo \& Keller 1996, 1997; Gandorfer 2000, 2002, 2005).

In the presence {of a magnetic field} the whole scattering process is perturbed and 
the ensuing polarization pattern altered.
This phenomenon, the Hanle effect, is important because it can be exploited to detect 
and measure magnetic fields in astrophysical plasmas and, in particular, 
in the solar or a stellar atmosphere 
(e.g., Stenflo 1991; Trujillo Bueno 2001; Casini \& Landi Degl'Innocenti 2007).
This goal requires a reliable modeling of the radiation field anisotropy
that generates the polarization pattern in the first place. 
The anisotropy is dominated by the local center-to-limb variation of the 
radiation field, but horizontal fluctuations of the physical properties in the atmosphere may alter it too.

In this paper we investigate the effect of horizontal fluctuations 
of the thermodynamic parameters on {scattering polarization},
how they compete with magnetic fields to modify the general polarization patterns {of spectral lines},
and how both effects may be disentangled for diagnostic purposes 
(see Manso Sainz \& Trujillo Bueno 1999 for a first attempt at this program).
We focus on weak opacity and Planck function fluctuations.
This allows a semi-analytical treatment of the problem which greatly simplifies 
the numerical analysis and allows solving two and three dimensional problems with 
the computational cost of a plane-parallel one.

We consider resonance scattering and the Hanle
effect in a two-level atom whose lower-level is unpolarized; 
{for definiteness, the results shown in this paper are for the $J_\ell=0\rightarrow J_u=1$ transition.}
We also consider Rayleigh and Thompson scattering 
by exploiting its formal similarity to the resonance scattering problem.
These problems can be treated within two alternative, but equivalent, formalisms.
First, we may note that the polarization state of the light
irradiating the collectivity of atoms, and the polarization state of the scattered
radiation are linearly related.
Therefore, the whole scattering process may be described through a {\em phase matrix}
relating the Stokes parameters of the incident and scattered beams; 
the source function of the radiative transfer equations is then an average over all 
possible incident angles of the phase matrix (e.g., Chandrasekhar 1960).
Alternatively, we may consider the intermediate state involving atoms explicitly.
The incident radiation excites the atoms, generating atomic level polarization (population imbalances and quantum coherences between sublevels); as a consequence, the reemited (scattered) radiation is linearly polarized.
A complete description of the excitation state of the atomic system taking into account
sublevel populations, coherences and statistical mixtures can be done through the 
density matrix (Fano 1957). 
Further, we will work with the spherical tensors components of the density matrix. 
In Sect.~\ref{sect2} we show how to formulate mathematically this problem using elements of this formalism.

The density matrix formalism provides several advantages for the problem under consideration.
First, its spherical tensor decomposition
is especially well-suited to apprehend and exploit the symmetries of the problem.
It will become apparent through the paper that symmetries lead to important 
simplifications in equations in a transparent way.
Second, it is computationally advantageous to pose the problem with the atomic density
matrix elements as unknowns, from which the radiation field straightforwardly derives,
rather than to consider the radiation field itself as the unknown.
This is because there are at most six density matrix components in our problem 
to be determined at each position in the medium, while 
there are three Stokes parameters that depend on the spatial, angular and frequency variables.
Finally, the density matrix is a fundamental and very general concept 
and polarization transfer theories based on it have been developed 
to treat more general problems than the ones considered here 
(Landi Degl'Innocenti \& Landolfi 2004), e.g., resonance scattering with lower-level atomic polarization, 
zero-field dichroism (Trujillo Bueno \& Landi Degl'Innocenti 1997; Trujillo Bueno 1999; 
Trujillo Bueno et al. 2002; Manso Sainz \& Trujillo Bueno 2003), and scattering in multilevel systems 
(Manso Sainz \& Landi Degl'Innocenti 2002; Manso Sainz, Landi Degl'Innocenti, \& Trujillo Bueno
2006; Manso Sainz \& Trujillo Bueno 2003, 2010; {{\v{S}t\v{e}p\'an} \& Trujillo Bueno 2010, 2011).
Therefore, the methods and techniques developed here can then be generalized 
to these other problems.

Coherent continuum scattering polarization, resonance scattering polarization, and the Hanle effect in  
plane-parallel media have been studied extensively.
On the contrary, in two and three dimensions, the problem has been attacked more sparingly
(but see Dittmann 1999; Paletou, Bommier, \& Faurobert-Scholl 1999; 
Manso Sainz \& Trujillo Bueno 1999; Manso Sainz 2002; Trujillo Bueno, Shchukina \& Asensio Ramos 2004;
Trujillo Bueno \& Schukina 2007, 2009; Schukina \& Trujillo Bueno 2011;
Anusha \& Nagendra 2011; Anusha, Nagendra, \& Paletou 2011).
This is because the inherent computational complexity of multidimensional
radiative transfer problems increases notably when polarization is
taken into account, due to its sensitivity to the radiation field angular dependence.
Here, we overcome these difficulties by considering scattering polarization transfer
in atmospheres that are only {\em weakly} inhomogeneous, in the sense that horizontal fluctuations
can be considered as a perturbation of a plane-parallel problem.
We linearize and derive equations for the fluctuations of the radiation field (Sect.~\ref{sect3}). 
Then, harmonic analysis of the linear problem is possible (Sect.~\ref{sect4}).
This approach allows application of simple symmetry arguments to simplify 
the problems and isolation of the relevant variables in each case, 
which leads to a system of transfer problems that are formally similar to 
the plane-parallel case and can thus be solved applying standard numerical techniques
(Sect.~\ref{sect5}).
This program generalizes the approach of Kneer \& Heasley (1979), Kneer (1981),
and Trujillo Bueno \& Kneer (1990) for the unpolarized problem.

\begin{figure}[t]\label{fig01}
\centerline{\epsfig{figure=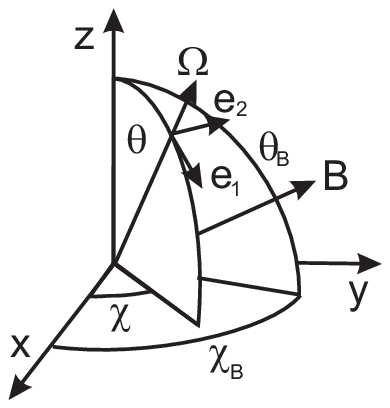, width=4cm}\epsfig{figure=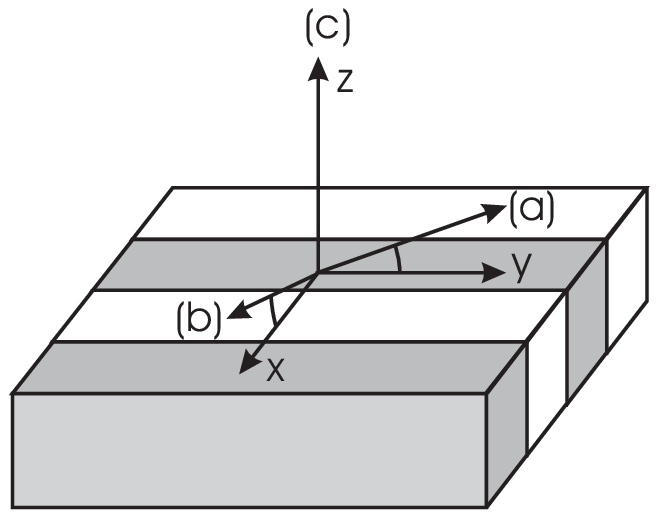, width=5.4cm}}
\caption{Left panel: reference system for polarization. 
A ray propagates along the direction
$\vec{\Omega}$, $\vec{e_1}$ is perpendicular to $\vec{\Omega}$ and lies on 
the meridian plane (the plane containing $\vec{\Omega}$ and the $z$-axis), and
$\vec{e_2}$ is perpendicular to both, $\vec{\Omega}$ and $\vec{e_1}$.
Note that $\vec{e_1}$, $\vec{e_2}$, $\vec{\Omega}$, form an orthonormal positive base.
In Eqs.~(\ref{eq02}) and (\ref{eq04}), the positive $Q$-direction
is taken to be along $\vec{e_1}$, i.e., perpendicular to the projected horizon (limb).
$\theta_B$ and $\chi_B$ are 
the polar and azimuthal angles of the magnetic field, respectively.
Right panel: three lines-of-sight (LOS) are of particular relevance in our 
analysis for two-dimensional media (invariant direction along the $y$-axis).
The slabs in the figure represent roughly the cosinusoidal fluctuactions considered in Sect.~4.
In case a, we observe close to the limb ($\mu=\cos \theta=0.1$) and along the invariant direction ($\chi=90^\circ$);
in case b, we observe close to the limb ($\mu=0.1$) and {\em accross the slabs} ($\chi=0^\circ$); in case c, we observe {\em at disk center} ($\mu=1$).
}
\end{figure}

\section{Formulation and general equations}\label{sect2}

This section states mathematically three polarization transfer problems:
resonance scattering (Sect. \ref{sect21}), coherent continuum scattering (Sect. \ref{sect22}), and the Hanle effect (Sect. \ref{sect23}) in optically thick atmospheres.
General symmetries of these problems mainly related to their dimensionality are naturally described within the formalism, and discussed in Sect. \ref{sect24}.

\subsection{Scattering line polarization}\label{sect21}

We consider resonance scattering polarization in an inhomogeneous, non-magnetized, static atmosphere.
In such a medium scattering produces linear polarization but not circular polarization.
Since the atmosphere is horizontally inhomogeneous, all three Stokes parameters
$I$, $Q$ and $U$ are necessary to describe the polarization state of the light.
The lower-level of the transition is assumed to be unpolarized (all its magnetic sublevels are equally populated). Therefore, there is no selective absorption of polarization components 
(i.e., no zero-field dichroism) as discussed in Trujillo Bueno \& Landi Degl'Innocenti (1997), Trujillo Bueno (1999) and Manso Sainz \& Trujillo Bueno (2003).
The transfer equations for polarization then read:
\begin{subequations}\label{eq01}
\begin{align}
  \frac{\rm d}{{\rm d}s}I&=-\kappa (I-S_I), 
                                                              \label{eq01a}
\displaybreak[0] \\
  \frac{\rm d}{{\rm d}s}Q&=-\kappa (Q-S_Q), 
                                                              \label{eq01b}
\displaybreak[0] \\
  \frac{\rm d}{{\rm d}s}U&=-\kappa (U-S_U), 
                                                              \label{eq01c}
\end{align}
\end{subequations}
where d$/$d$s$ is the derivative along the ray path, $\kappa$ is the absorption
coefficient, and $S_I$, $S_Q$ and $S_U$ the source functions for the corresponding
Stokes parameters.
In the presence of a background continuum the absorption coefficient
has contributions from the line and the continuum: 
$\kappa=\kappa^{\rm line}\phi_\nu+\kappa^{\rm cont}$ ($\phi_\nu$ is the line
absorption profile), while the source function 
$S_I=r_\nu S^{\rm line}_I + (1-r_\nu) S^{\rm cont}_I$, with 
$r_\nu={\kappa^{\rm line}\phi_\nu}/{\kappa}$. 
We will assume the continuum to be in LTE, $S^{\rm cont}_I=B_\nu(T)$, ($B_\nu(T)$ is the Planck function) 
and unpolarized ($S_Q=r_\nu S^{\rm line}_Q$, $S_U=r_\nu S^{\rm line}_U$).
A polarized continuum due to Rayleigh or Thompson scattering
can be readily included without difficulty (see next subsection).

The line source functions depend on the excitation state of the atoms which,
within the density matrix formalism, is described through the density matrix $\rho$ (Fano 1957; Blum 1981).
Its diagonal elements $\rho_{mm}$ are the populations of the individual magnetic sublevels 
and the non-diagonal ones the coherences between them.
It is convenient to express the density matrix in its spherical components $\rho^K_Q$ 
($K=0, ..., 2J$, $Q=-K, ..., K$; Brink \& Satchler 1968).
Introducing the new variables 
$S^K_Q=\frac{2h\nu^3}{c^2}\frac{2J_\ell +1}{\sqrt{2J_u+1}}\rho^K_Q/{\cal N}_\ell$
($\nu$ is the frequency of the transition and ${\cal N}_\ell$ the total population 
of the ground level) we will derive expressions that are a natural generalization
of the standard radiative transfer theory for unpolarized radiation.
The $S^K_Q$ components are, in general, complex quantities satisfying 
the conjugation property $[S^K_{Q}]^*=(-1)^QS^K_{-Q}$ 
(``*'' stands for complex conjugation), as a consequence of the hermiticity 
of the density matrix $\rho$.
For computational purposes it is more convenient to work with the following real quantities 
\begin{align}
\tilde{S}^K_Q&=\frac{1}{2}[S^K_Q+(-1)^QS^K_{-Q}], \label{eq02_2}
\\
\hat{S}^K_Q&=\frac{1}{2{\rm i}}[S^K_Q-(-1)^QS^K_{-Q}],\label{eq03_2}
\end{align}
defined for $Q>0$ (the equivalent $Q<0$ elements are redundant).
Note that $\tilde{S}^K_Q$ and $\hat{S}^K_Q$ are the real and imaginary
parts of $S^K_Q$ ($Q>0$).

The line source functions can be obtained applying the density matrix theory of polarization in spectral lines (e.g., Landi Degl'Innocenti \& Landolfi 2004) assuming complete frequency redistribution.
They can be expressed as
\begin{subequations}\label{eq02}
\begin{align}
  \begin{split}
  S^{\rm line}_I&=\,S^0_0 + w_{J_uJ_l}^{(2)} \Big{\{}
  \frac{1}{2\sqrt{2}} (3 \mu^2-1)S^2_0 - \sqrt{3} \mu \sqrt{1-\mu^2} 
  (\cos \chi \tilde{S}^2_1 - \sin \chi \hat{S}^2_1) \\ 
  &\quad+ \frac{\sqrt{3}}{2} (1-\mu^2) (\cos 2\chi \, \tilde{S}^2_2
  -\sin 2\chi \, \hat{S}^2_2) \Big{\}},
  \end{split}                                                  
                                                              \label{eq02a}
\displaybreak[0] \\
  \begin{split}
  S^{\rm line}_Q&=\,w_{J_uJ_l}^{(2)}\Big{\{} \frac{3}{2\sqrt{2}}(\mu^2-1) S^2_0 -
  \sqrt{3}  \mu \sqrt{1-\mu^2} (\cos \chi
  \tilde{S}^2_1 - \sin \chi \hat{S}^2_1) \\ 
  &\quad- \frac{\sqrt{3}}{2} (1+\mu^2) (\cos 2\chi \, \tilde{S}^2_2
  -\sin 2\chi \, \hat{S}^2_2)\Big{\}},   
  \end{split}
                                                              \label{eq02b}
\displaybreak[0] \\
  \begin{split}
  S^{\rm line}_U&=\, w_{J_uJ_l}^{(2)}\sqrt{3} \Big{\{} \sqrt{1-\mu^2} ( \sin \chi
  \tilde{S}^2_1+\cos \chi \hat{S}^2_1) \\ &\quad+
  \mu (\sin 2\chi \, \tilde{S}^2_2 + \cos 2\chi \, \hat{S}^2_2)\Big{\}}.
  \end{split}
                                                              \label{eq02c}
\end{align}
\end{subequations}
where $\theta=\arccos\mu$ and $\chi$ are the inclination and 
azimuth of the ray, respectively, and the reference
directions for polarization are shown in Fig.~\ref{fig01} 
(i.e., the positive-$Q$ direction is along $\vec{e_1}$). 
The coefficient $w_{J_uJ_\ell}^{(2)}$ is a numerical
factor that only depends on the quantum numbers of the transition considered 
(see Eq.~(10.12) in Landi Degl'Innocenti \& Landolfi 2004). 
Explicit values for several transitions, as well as the limits for large $J$,
are compiled in Table 1.

\begin{table}[t]
\caption{$w_{J_uJ_\ell}^{(2)}$}
\begin{tabular}{cccccccccccccc}
\hline\\[-4pt]
 & \multicolumn{13}{c}{$J_u$}  \\[4pt]
$J_\ell$ & 1 & 3/2 & 2 & 5/2 & 3 & 7/2 & 4 & 9/2 & 5 & 11/2 & 6 & 13/2& $\infty$ \\[4pt]
\hline\\[-4pt]
$J_u-1$ & 1 & $\frac{1}{\sqrt{2}}$ & $\frac{1}{2}\sqrt{\frac{7}{5}}$ & $\frac{\sqrt{7}}{5}$ & $\frac{\sqrt{6}}{5}$ & $\frac{\sqrt{42}}{14}$ & $\frac{\sqrt{154}}{28}$ & $\frac{1}{2}\sqrt{\frac{11}{15}}$ & $\frac{1}{5}\sqrt{\frac{13}{3}}$ & $\frac{1}{5}\sqrt{\frac{91}{22}}$ & $\frac{1}{2}\sqrt{\frac{7}{11}}$ & $\sqrt{\frac{2}{13}}$  & $\frac{1}{\sqrt{10}}$ \\[7pt]
$J_u$   & $-\frac{1}{2}$ & $-\frac{2\sqrt{2}}{5}$ & $-\frac{1}{2}\sqrt{\frac{7}{5}}$ & $-\frac{8}{5\sqrt{7}}$ & $-\frac{1}{2}\sqrt{\frac{3}{2}}$ & $-2\sqrt{\frac{2}{21}}$ & $-10\sqrt{\frac{77}{2}}$ & $-\frac{8}{\sqrt{165}}$ & $-\frac{\sqrt{39}}{10}$ & $-2\sqrt{\frac{14}{143}}$ & $-\frac{1}{2}\sqrt{\frac{11}{7}}$ & $-\frac{8}{5}\sqrt{\frac{2}{13}}$ & $-\sqrt{\frac{2}{5}}$ \\[7pt]
$J_u+1$ & $\frac{1}{10}$ & $\frac{1}{5\sqrt{2}}$ & $\frac{1}{\sqrt{35}}$ & $\frac{1}{2\sqrt{7}}$ & $\frac{1}{2\sqrt{6}}$ & $\frac{1}{5}\sqrt{\frac{7}{6}}$ & $\frac{1}{5}\sqrt{\frac{14}{11}}$ & $\sqrt{\frac{3}{55}}$ & $\frac{1}{2}\sqrt{\frac{3}{13}}$ & $\sqrt{\frac{11}{182}}$ & $\frac{1}{5}\sqrt{\frac{11}{7}}$ & $\frac{1}{10}\sqrt{\frac{13}{2}}$ & $\frac{1}{\sqrt{10}}$ \\[7pt]
\hline\\[-4pt]
\end{tabular}
\end{table}

Equations (\ref{eq02}) express the angle-dependent line source functions in terms
of the six (angle and frequency independent) variables 
$S^0_0$, $S^2_0$, $\tilde{S}^2_1$, $\hat{S}^2_1$, $\tilde{S}^2_2$, $\hat{S}^2_2$ 
satisfying (e.g., Trujillo Bueno \& Manso Sainz 1999):
\begin{subequations}\label{eq03}
\begin{align}
  S^0_0 &=\,(1-\epsilon) J^0_0 \,+\, \epsilon B_{\nu_{ul}},
                                                              \label{eq03a}
\displaybreak[0]\\
  [1+\delta^{(2)}(1-\epsilon)] S^2_0 &=\,(1-\epsilon)w_{J_uJ_l}^{(2)} J^2_0, 
                                                              \label{eq03b}
\displaybreak[0]\\
  [1+\delta^{(2)}(1-\epsilon)] \tilde{S}^2_1 &=\,(1-\epsilon)w_{J_uJ_l}^{(2)} \tilde{J}^2_1, 
                                                              \label{eq03c}
\displaybreak[0]\\
  [1+\delta^{(2)}(1-\epsilon)] \hat{S}^2_1 &=\,-(1-\epsilon)w_{J_uJ_l}^{(2)} \hat{J}^2_1,        
                                                              \label{eq03d}
\displaybreak[0]\\
  [1+\delta^{(2)}(1-\epsilon)] \tilde{S}^2_2 &=\,(1-\epsilon)w_{J_uJ_l}^{(2)} \tilde{J}^2_2,      
                                                              \label{eq03e}
\displaybreak[0]\\
  [1+\delta^{(2)}(1-\epsilon)] \hat{S}^2_2 &=\,-(1-\epsilon)w_{J_uJ_l}^{(2)} \hat{J}^2_2.   
                                                              \label{eq03f}
\end{align}
\end{subequations}
In Eqs.~(\ref{eq03}), $\epsilon=C_{u\ell}/(C_{u\ell}+A_{u\ell})$
is the collisional destruction probability ($C_{u\ell}$ is the 
collisional desexcitation rate and $A_{u\ell}$ the Einstein coefficient for 
spontaneous emission), and $\delta^{(2)}=D^{(2)}/A_{u\ell}$ is the depolarizing 
rate due to elastic collisions with neutral hydrogen normalized to $A_{u\ell}$. 
Finally, the radiation field tensor components read:
\begin{subequations}\label{eq04}
\begin{align}
  J^0_0(\nu)&=  \oint \frac{{\rm d}
    \vec{\Omega'}}{4\pi}\,I_{\nu \vec{\Omega'}},                          
                                                              \label{eq04a}
\displaybreak[0]\\
  J^2_0(\nu)&= \oint \frac{{\rm d} \vec{\Omega'}}{4\pi}
  \frac{1}{2\sqrt{2}} [(3\mu'^2-1)I_{\nu \vec{\Omega'}}+3(\mu'^2-1)Q_{\nu
  \vec{\Omega'}}],
                                                              \label{eq04b}
\displaybreak[0]\\
  \tilde{J}^2_1(\nu)&= \oint \frac{{\rm d} \vec{\Omega'}}{4\pi}
  \frac{\sqrt{3}}{2} \sqrt{1-\mu'^2}[-\mu'\cos\chi'(I_{\nu \vec{\Omega'}}+Q_{\nu
  \vec{\Omega'}})+\sin\chi' U_{\nu \vec{\Omega'}}],
                                                              \label{eq04c}
\displaybreak[0]\\
  \hat{J}^2_1(\nu) &= \oint \frac{{\rm d} \vec{\Omega'}}{4\pi}
  \frac{\sqrt{3}}{2} \sqrt{1-\mu'^2}[-\mu'\sin\chi'(I_{\nu \vec{\Omega'}}+Q_{\nu
  \vec{\Omega'}})-\cos\chi' U_{\nu \vec{\Omega'}}],
                                                              \label{eq04d}
\displaybreak[0]\\
  \tilde{J}^2_2(\nu) &= \oint \frac{{\rm d} \vec{\Omega'}}{4\pi}
  \frac{\sqrt{3}}{4} \,
  [\cos(2\chi')[(1-\mu'^2)I_{\nu \vec{\Omega'}}-(1+\mu'^2)Q_{\nu
  \vec{\Omega'}}]+2\sin(2\chi')\mu' U_{\nu \vec{\Omega'}}],
                                                              \label{eq04e}
\displaybreak[0]\\
  \hat{J}^2_2(\nu) &= \oint \frac{{\rm d} \vec{\Omega'}}{4\pi}
  \frac{\sqrt{3}}{4} \,
  [\sin(2\chi')[(1-\mu'^2)I_{\nu \vec{\Omega'}}-(1+\mu'^2)Q_{\nu
  \vec{\Omega'}}]-2\cos(2\chi')\mu' U_{\nu \vec{\Omega'}}],
                                                              \label{eq04f}
\end{align}
\end{subequations}
where ${\rm d}\vec{\Omega'}=\sin\theta'{\rm d}\theta'{\rm d}\chi'={\rm d}\mu'{\rm d}\chi'$.
The $\tilde{J}^2_Q$ and $\hat{J}^2_Q$ components with $Q=1, 2$, in Eqs.~(\ref{eq04}) correspond
to the real and imaginary parts, respectively, of the complex spherical 
component $J^2_Q$ of the radiation tensors discussed in 
Sect.~5.1 of Landi Degl'Innocenti \& Landolfi (2004). 
Assuming complete frequency redistribution, 
the actual $J^K_Q$ components to be introduced in Eqs.~(\ref{eq03}) 
are the frequency averages over the absorption profile:
\begin{equation}\label{eq04_2}
J^K_Q=\int {\rm d}\nu \,\phi_\nu J^K_Q(\nu).
\end{equation}
Hereafter, we will suppress the subscripts expressing explicitly the dependence
of the different variables on $\nu$, unless necessary to avoid ambiguities.

In the limit $\delta^{(2)}\rightarrow\infty$, the $S^2_Q$ elements vanish
(see Eqs.~(\ref{eq03b})-(\ref{eq03f})). Then $S^{\rm line}_Q=S^{\rm line}_U=0$, 
$S^{\rm line}_I=S^0_0$ (Eqs.~(\ref{eq02})), and 
we recover the well-known expression for a two level atom 
under complete frequency redistribution (Eq.~{\ref{eq03a}}) 
of the classical radiative transfer theory for unpolarized radiation (e.g., Mihalas 1978).

We note in passing, that the equivalent phase matrix formalism is recovered 
by writing explicitly the expressions for the radiation field tensors 
(Eqs.~(\ref{eq04})-(\ref{eq04_2}))
into Eqs.~(\ref{eq03}), and then into the expressions for the source functions (Eqs.~(\ref{eq02})).
Thus proceeding we obtain explicit expressions for the source functions of the Stokes parameters,
at a given angle and frequency, as averages over the incident radiation field.
The linear operator relating incident and emergent radiation is the so-called
phase matrix (e.g., Chandrasekhar 1960).
The methods developed here are thus straightforwardly extended to the phase matrix formalism.

\subsection{Coherent continuum scattering polarization}\label{sect22}

Scattering by free electrons (Thompson scattering) and by atoms (most notably, neutral hydrogen) in the far wings of resonance lines (Rayleigh scattering) is coherent in the electron/atomic rest frame (Dirac 1925), but otherwise, the scattering phase matrix is equivalent to that of resonance scattering in a $J_\ell=0\rightarrow J_u=1$ transition (e.g., Chandrasekhar 1960).
Therefore,  we may write $w_{10}^{(2)}=1$ in Eqs.~(\ref{eq02}), $\epsilon=\delta^{(2)}=0$ in
Eqs.~(\ref{eq03}), and consider Eqs.~(\ref{eq04}) as a convenient factorization of the phase matrix in terms of six variables $J^K_Q$ that {are angle independent.}

For passing to the laboratory frame, we must consider the Doppler effect of the Maxwellian distribution of the scatterers velocities.
This implies convolving the incident radiation field, at each frequency $\nu'$ and angle $\mu'$, with a profile 
\begin{equation}\label{eq06}
\phi(\Theta; \nu, \nu')=\frac{1}{\Delta_D\sqrt{2\pi(1-\cos\Theta)}}\exp\left\{-\frac{(\nu-\nu')^2}{2(1-\cos\Theta)\Delta_D^2}\right\},
\end{equation}
where $\Theta=\arccos(\sqrt{1-\mu^2}\sqrt{1-\mu'^2}\cos(\chi-\chi')+\mu\mu')$ is the angle between incident and scattered beams, and $\Delta_D$ the Doppler width corresponding to the scatterers (note, in passing, that $\Delta_D({\rm e}^-)\gg\Delta_D({\rm H}^\circ)$).

If the incident continuum radiation field is spectrally flat, a convolution over $\nu'$ with the expression in Eq.~(\ref{eq06}) leaves the spectrum unaltered. 
Scattering can thus be considered coherent also in the laboratory frame, and
continuum scattering is described by exactly the same equations of the previous subsection 
with the formal substitutions $w_{10}^{(2)}=1$, $\epsilon=\delta^{(2)}=0$, $\phi_\nu=\delta(\nu-\nu')$.
Such an approximation for the continuum is formally equivalent 
to the complete frequency redistribution approximation for line scattering
by atoms. 

It is important to recall
that this approximation neglects important effects arising from the presence of spectral structure in the incident radiation field.
In particular, such an approximation should be considered with care (if at all)
when interpreting the continuum in the vicinity of spectral lines 
(see Landi Degl'Innocenti \& Landolfi 2004).

The source functions $S_I^{\rm scatt}$, $S_Q^{\rm scatt}$ and $S_U^{\rm scatt}$ 
due to pure coherent Rayleigh or Thompson
scattering in the continuum are given by Eqs.~(\ref{eq02})-(\ref{eq03})
with the following formal substitutions: $w^{(2)}_{J_uJ_\ell}=1$, $\epsilon=\delta^{(2)}=0$. 
Let $\sigma$ be the absorption coefficient for scattering, $\kappa^{\rm cont}$
the {\em true} continuum absorption coefficient due to thermal absorption,
$\kappa=\sigma+\kappa^{\rm cont}$, and $s=\sigma/\kappa$,
then the total source functions are $S_I=sS_I^{\rm scatt}+(1-s)B_\nu$,
$S_Q=sS_Q^{\rm scatt}$, and $S_U=sS_U^{\rm scatt}$.

\subsection{Hanle effect}\label{sect23}

The Hanle effect is the modification of scattering line polarization due to the presence 
of weak magnetic fields.
By {\em weak} magnetic field we mean the regime in which the splitting of the emission and absorption 
profiles ($\sim\nu_{\rm L}$, the Larmor frequency), is of the order of the natural width of the upper level of the transition (i.e., $A_{u\ell}$\footnote{Note that we are  assuming that the lower level is unpolarized and the so-called lower-level Hanle effect regime is not considered in this paper.}), and hence, negligible compared to the Doppler line width $\Delta\nu_{\rm D}$.
In this limit, the splitting of the line profile in its $\sigma$ and $\pi$ components is negligible
and the only noticeable effect of the magnetic field on the polarization state of 
light stems from its perturbing the excitation and coherence state of the atoms.
Therefore, the radiative transfer equations are still given by Eqs.~(\ref{eq01}) and (\ref{eq02}).
However, the excitation state of the atomic system is now a competition between radiative
excitation and partial decoherence induced by the magnetic field splitting (besides 
collisional excitation/relaxation).
The statistical equilibrium equations for the $S^K_Q$ components read (e.g., Manso Sainz \& Trujillo Bueno 1999)
\begin{subequations}\label{eq10}
\begin{align}
  S^0_0 &=\,\;\;(1-\epsilon) J^0_0 \,+\, \epsilon B_{\nu_{ul}},
                                                              \label{eq10a}
\displaybreak[0]\\
  [1+\delta^{(2)}(1-\epsilon)] S^2_0 &=\,\;\;(1-\epsilon)w^{(2)}_{J_uJ_l} J^2_0
  -\Gamma(1-\epsilon) \sqrt{6}s_{\theta_B}(s_{\chi_B}
  \tilde{S}^2_1 + c_{\chi_B} \hat{S}^2_1), 
                                                              \label{eq10b}
\displaybreak[0]\\
  [1+\delta^{(2)}(1-\epsilon)] \tilde{S}^2_1 &=\,\;\;(1-\epsilon)w^{(2)}_{J_uJ_l} \tilde{J}^2_1
-\Gamma(1-\epsilon) (-\sqrt{\frac{3}{2}}
s_{\theta_B}s_{\chi_B} S^2_0 + c_{\theta_B} \hat{S}^2_1 + s_{\theta_B}s_{\chi_B} \tilde{S}^2_2 + 
s_{\theta_B}c_{\chi_B}\hat{S}^2_2), 
                                                              \label{eq10c}
\displaybreak[0]\\
  [1+\delta^{(2)}(1-\epsilon)] \hat{S}^2_1 &=-(1-\epsilon)w^{(2)}_{J_uJ_l} \hat{J}^2_1
  -\Gamma(1-\epsilon) (-\sqrt{\frac{3}{2}} s_{\theta_B}c_{\chi_B}S^2_0
  - c_{\theta_B} \tilde{S}^2_1 -s_{\theta_B}c_{\chi_B} \tilde{S}^2_2 + s_{\theta_B}s_{\chi_B}
  \hat{S}^2_2),        
                                                              \label{eq10d}
\displaybreak[0]\\
  [1+\delta^{(2)}(1-\epsilon)] \tilde{S}^2_2 &=\,\;\;(1-\epsilon)w^{(2)}_{J_uJ_l} \tilde{J}^2_2 
  -\Gamma(1-\epsilon) (-s_{\theta_B}s_{\chi_B} \tilde{S}^2_1 
  + s_{\theta_B}c_{\chi_B}  \hat{S}^2_1 + 2c_{\theta_B}\hat{S}^2_2),
                                                              \label{eq10e}
\displaybreak[0]\\
  [1+\delta^{(2)}(1-\epsilon)] \hat{S}^2_2 &=-(1-\epsilon)w^{(2)}_{J_uJ_l} \hat{J}^2_2
  -\Gamma(1-\epsilon) (-s_{\theta_B}c_{\chi_B} \tilde{S}^2_1 - s_{\theta_B}s_{\chi_B}\hat{S}^2_1 
  -2c_{\theta_B} \tilde{S}^2_2).
                                                              \label{eq10f}
\end{align}
\end{subequations}
In Eqs.~(\ref{eq10b})-(\ref{eq10f}), $\Gamma=0.879\times 10^7 g_{u}B/A_{u\ell}$ 
($g_{u}$ is the upper-level Land\'e factor and $B$ the magnetic field strength in gauss), 
$s_\alpha=\sin\alpha$, $c_\alpha=\cos\alpha$, and
$\theta_B$ and $\chi_B$ are the inclination and azimuth of the magnetic field, respectively
(see Fig.~1).
 
The micro-turbulent limit of these equations is often studied because
it leads to simpler equations and it is considered a suitable approximation  
to interpret spectropolarimetric observations with low spatio-temporal resolution 
(Stenflo 1982, 1994; Landi Degl'Innocenti 1985).
In this limit, one assumes that the magnetic field changes its orientation and 
strength at scales much smaller than the photon mean free path.
The excitation state of the atom at a point in the atmosphere
is then the average over all possible realizations of the magnetic field and
the $S^K_Q$ components satisfy averaged Eqs.~(\ref{eq10}) (once formally solved).
Averaging over an isotropic distribution of fields of constant strength 
gives equations identical to those of the pure scattering case (Eqs.~(\ref{eq03})), 
but with the $J^2_Q$ radiation field tensor components multiplied by a simple numerical factor
(Trujillo Bueno \& Manso Sainz 1999)
\begin{equation}\label{eq11}
{\cal H}^{(2)}=\frac{1}{5}\left(1+\frac{2}{1+\gamma^2}+\frac{2}{1+4\gamma^2}\right),
\end{equation}
where $\gamma=\Gamma(1-\epsilon)/[1+\delta^{(2)}(1-\epsilon)]$.

Note that the same symmetry considerations given for scattering polarization in 
Sect.~\ref{sect21} hold in the presence of an isotropic micro-turbulent field.

\subsection{Symmetry considerations}\label{sect24}

Equations (\ref{eq01}), together with Eqs.~(\ref{eq02})-(\ref{eq04}) are general and describe 
the transfer of polarized line radiation in a scattering atmosphere of
arbitrary geometry. 
However, the presence of symmetries in the medium leads to important simplifications.
Thus, in a plane-parallel atmosphere, symmetry imposes that light be linearly polarized 
either parallel to the horizon or perpendicular to it. Choosing these directions
as the reference for polarization (see Fig.~(\ref{fig01})), the only non-vanishing Stokes
parameters are $I$ and $Q$, and they are independent of the azimuthal angle.
Consequently, $J^2_Q=0$ for $Q\ne 0$, which implies 
$S^2_{Q\ne 0}=0$, and we recover the well-known
expressions for scattering line polarization in a plane-parallel medium
(e.g., Trujillo Bueno \& Manso Sainz 1999).

Consider a general Cartesian two-dimensional medium. We choose without loss of generality
the $y$-axis as the invariant direction. Then, the following symmetries hold
(see Fig.~\ref{fig02} for proof):
\begin{subequations}\label{eq05}
\begin{align}
I_{\mu, \chi} &= I_{\mu, -\chi},\label{eq05a} \\
Q_{\mu, \chi} &= Q_{\mu, -\chi},\label{eq05b} \\
U_{\mu, \chi} &= -U_{\mu, -\chi}. \label{eq05c}
\end{align}
\end{subequations}
Introducing these expressions into Eqs.~(\ref{eq04}), one finds that  
$\hat{J}^2_1=\hat{J}^2_2=0$. Therefore, $\hat{S}^2_1=\hat{S}^2_2=0$, 
and just four variables ($S^0_0$, $S^2_0$, $\tilde{S}^2_1$, 
and $\tilde{S}^2_2$) are necessary to completely describe the excitation 
state of the atomic system in a two-dimensional medium.

The symmetries of the problem can be further exploited, but this requires 
a more explicit knowledge of the horizontal fluctuations of the physical system (Sect.~\ref{sect4}).
Clearly, the same symmetry considerations  apply to the coherent continuum scattering polarization case as well.

\begin{figure}[t]
\centerline{\epsfig{figure=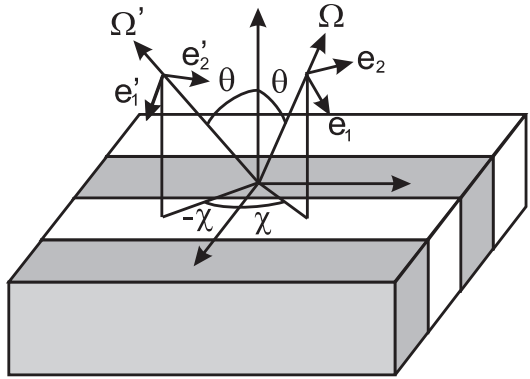, width=7cm}}
\caption{A two-dimensional scattering medium is invariant under 
translations along a direction (say, the $y$-axis), and 
reflections on the $x$-$z$ plane. 
Thus, at an arbitrary point in the atmosphere the polarization 
of light propagating along
two rays $\mvec{\Omega}$ and $\vec{\Omega'}$ is 
obtained by reflection from the other. 
Light polarized along directions $\vec{e_1}$ and 
$\vec{e_2}$ in ray $\mvec{\Omega}$, is polarized along directions
$\vec{e_1'}$ and $\vec{e_2'}$, respectively, in ray $\mvec{\Omega}'$.
Hence, $I_{\mu, \chi}= I_{\mu, -\chi}$ and $Q_{\mu, \chi} = Q_{\mu, -\chi}$
(Eqs.~(\ref{eq05a})-(\ref{eq05b})). 
However, light polarized along the direction
$(\vec{e_1}\pm\vec{e_2})/\sqrt{2}$ in ray $\mvec{\Omega}$ is polarized 
along direction $(\vec{e_1'}\mp\vec{e_2'})/\sqrt{2}$ in ray $\mvec{\Omega'}$ 
and $U_{\mu, \chi} = -U_{\mu, -\chi}$ (Eq.~(\ref{eq05c})).
Note that, therefore, $U\equiv 0$ at right angles from the fluctuations ($\chi=0$) and, in particular, at disk center observation ($\mu=1$).\label{fig02}}
\end{figure}

In general, the presence of a deterministic magnetic field breaks the symmetry of
the radiation field. Thus, regardless of the problem's dimensionality three Stokes
parameters ($I$, $Q$, and $U$) are necessary to describe the polarization state 
of the radiation field, which is no longer axially symmetric. Therefore, the $J^2_Q$
elements do not vanish in general and all six $S^K_Q$ elements are necessary 
to describe the excitation state of the atomic system.
The description simplifies only for very special geometrical configurations.
Thus, if the medium is planeparallel and the magnetic field vertical, we recover the
symmetries of the scattering polarization in a planeparallel medium  
discussed in Sect.~\ref{sect21}.
In fact, that problem is completely equivalent to the scattering polarization problem: {\em the Hanle effect
does not operate}.

In a two-dimensional medium with no magnetic field, just four $J^K_Q$ ($S^K_Q$) components 
are necessary  to describe the illumination (excitation state) in the medium 
(Sect.~\ref{sect21}; Manso Sainz 2002). This holds true also in the presence of a horizontal magnetic field
aligned along the invariant direction since in that case the symmetries illustrated 
in Fig.~\ref{fig02} are also satisfied.
Note, however, that unlike the planeparallel case, a vertical magnetic field 
does modify scattering polarization and hence the Hanle effect does not vanish.

\section{Weak fluctuations and linearization}\label{sect3}

\subsection{Radiative transfer equations}\label{sect31}

Now, we consider a weakly inhomogeneous atmosphere along the horizontal direction.
We assume that all the physical variables can be expressed as a small perturbation of a 
plane-parallel atmosphere and that we can thus neglect {\em second order} terms. 
We write
\begin{align}
\kappa(x, y, z) &= \bar{\kappa}(z) + \delta \kappa(x, y, z), \label{eq13} \\
S_I(x, y, z) &= \bar{S_I}(z) + \delta S_I(x, y, z), \label{eq14} \\
I(x, y, z) &= \bar{I}(z) + \delta I(x, y, z), \label{eq15}
\end{align}
(with analogous expressions for $Q$ and $U$), 
where $\bar{I}$ is the solution of the planeparallel transfer problem posed by
$\bar{\kappa}$ and $\bar{S}_I$. (To avoid lengthy and heavy algebraic expressions,
throughout the paper we will drop explicit dependencies
on the independent variables ---space, frequency, angles--- whenever they
are not strictly necessary and does not lead to ambiguities.)
Introducing expressions (\ref{eq13})-(\ref{eq15}) 
into the radiative transfer equations (\ref{eq01}) we get
\begin{subequations}\label{eq16}
\begin{align}
  \frac{\rm d}{{\rm d} s} \delta I
	&=-\bar{\kappa} (\delta I-\delta S_I)
	-\delta\kappa 
	(\bar{I} + \delta I 
	- \bar{S}_I - \delta S_I),      
                                                              \label{eq16a}
\displaybreak[0] \\
  \frac{\rm d}{{\rm d} s} \delta Q
	&=-\bar{\kappa} (\delta Q-\delta S_Q)
	-\delta\kappa 
	(\bar{Q} + \delta Q 
	- \bar{S}_Q - \delta S_Q),      
                                                              \label{eq16b}
\displaybreak[0] \\
  \frac{\rm d}{{\rm d} s} \delta U
	&=-\bar{\kappa} (\delta U-\delta S_U)
	-\delta\kappa 
	(\bar{U} + \delta U 
	- \bar{S}_U - \delta S_U).
                                                              \label{eq16c}
\end{align}
\end{subequations}
If $\delta \kappa=0$, Eqs.~(\ref{eq16}) are linear, irrespectively
of the magnitude of the amplitude of the source function fluctuation.
In the presence of small opacity perturbations ($\delta \kappa/\bar{\kappa}\ll 1$), and
assuming that the ensuing perturbation on the intensity and linear polarization is small 
($\delta I/\bar{I}, \delta Q/\bar{I}, \delta U/\bar{I} \ll 1$), we 
neglect second order terms in Eqs.~(\ref{eq16}):
\begin{subequations}\label{eq17}
\begin{align}
  \frac{\rm d}{{\rm d} s} \delta I
	&=-\bar{\kappa} (\delta I-\delta S_I^{\rm eff}),
                                                              \label{eq17a}
\displaybreak[0] \\
  \frac{\rm d}{{\rm d} s} \delta Q
	&=-\bar{\kappa} (\delta Q-\delta S_Q^{\rm eff}),
                                                              \label{eq17b}
\displaybreak[0] \\
  \frac{\rm d}{{\rm d} s} \delta U
	&=-\bar{\kappa} (\delta U-\delta S_U^{\rm eff}),
                                                              \label{eq17c}
\end{align}
\end{subequations}
where the {\em effective} source functions are
\begin{subequations}\label{eq18}
\begin{align}
   \delta S^{\rm eff}_I&=\, \delta S_I - \frac{\delta\kappa}
	{\bar{\kappa}}(\bar{I} - \bar{S}_I), 
                                                              \label{eq18a}
\displaybreak[0] \\
   \delta S^{\rm eff}_Q&=\, \delta S_Q - \frac{\delta\kappa}
	{\bar{\kappa}}(\bar{Q} - \bar{S}_Q), 
                                                              \label{eq18b}
\displaybreak[0] \\
   \delta S^{\rm eff}_U&=\, \delta S_U - \frac{\delta\kappa}
	{\bar{\kappa}}(\bar{U} - \bar{S}_U). 
                                                              \label{eq18c}
\end{align}
\end{subequations}
Equations (\ref{eq17})-(\ref{eq18}) are linear in the sense that if $\delta_1 I$ and $\delta_2 I$
are solutions of transfer problems with $\delta_1\kappa$, $\delta_1 S_I$, and
$\delta_2\kappa$, $\delta_2S_I$, respectively, then 
$\alpha\delta_1 I+\beta\delta_2 I$ is the solution of the transfer problem with
fluctuations $\alpha\delta_1\kappa+\beta\delta_2\kappa$ and
$\alpha\delta_1S_I+\beta\delta_2S_I$, and analogously for $Q$ and $U$.
Note in passing, that due to linearity, a problem with fluctuations on both, opacity and source 
function, can be retrieved from the superposition of a problem in which only
the opacity fluctuates (while the source function remains constant), and the problem
in which only the source function fluctuates (while the opacity is fixed).

We will consider the following boundary conditions. An open boundary
with no incident illumination at the top ($z=z_{M}$): $\delta I_{\mu\lambda}(x, y, z_M)=\delta Q_{\mu\lambda}(x, y, z_M)=\delta U_{\mu\lambda}(x, y, z_M) = 0$, 
for $\mu<0$; 
a thermal, unpolarized radiation field at the bottom ($z=z_{m}$), 
well below the thermalization depth: $\delta I_{\mu\lambda}(x, y, z_m) = \delta B_\nu$, $\delta Q_{\mu\lambda}(x, y, z_m) =\delta U_{\mu\lambda}(x, y, z_m) = 0$
for $\mu>0$.

\subsection{Statistical equilibrium equations}\label{sect32}

The statistical equilibrium equations for all the problems considered in the previous section can be expressed as an algebraic system of equations:
\begin{equation}
M\mvec{S}=\mvec{b},
\end{equation}
where $\mvec{S}$ is a formal vector whose (up to 6) elements are the $S^K_Q$ components, 
$M$ is a square matrix whose coefficients depend on the parameters of the problem (magnetic field strength and orientation, collisional rates $\epsilon$, $\delta$), and the components of the formal vector $\mvec{b}$ are linear combinations of $J^K_Q$ components and the Planck function. 

Here we consider fluctuations in the Planck function and opacity:
\begin{eqnarray}\label{eq20}
B_\nu&=&\bar{B}_\nu+\delta B_\nu, \\
\kappa&=&\bar{\kappa}+\delta \kappa.
\end{eqnarray}
This implies fluctuations for $S^K_Q$ and hence, for the radiation field tensors, which are decomposed as
\begin{equation}\label{eq19bis}
S^K_Q=\bar{S}^K_Q+\delta S^K_Q,
\end{equation}
\begin{equation}\label{eq19}
J^K_Q=\bar{J}^K_Q+\delta J^K_Q.
\end{equation}
We thus find that if the mean unperturbed variables satisfy the statistical equilibrium equations $M\mvec{\bar{S}}=\mvec{\bar{b}}$, the perturbations satisfy the same system $M\mvec{\delta{S}}=\mvec{\delta{b}}$.

It is possible to consider more general fluctuations in other physical parameters of the problem, as for example, in the magnetic field or in the collisional rates. 
It is then possible to linearize the statistical equilibrium equations which yields the 
following formal equations for the perturbations
\begin{equation}
M\mvec{\delta{S}}=\mvec{\delta{b}}-\delta M\mvec{\bar{S}}.
\end{equation}



\section{Harmonic analysis}\label{sect4}

\subsection{Radiative transfer equation}\label{sect41}

We perform an harmonic analysis of the linear problem posed in the previous section and
we show that the general multidimensional 
radiative transfer problem posed by Eqs.~(\ref{eq17})-(\ref{eq18})
reduces to several coupled plane-parallel radiative transfer problems.
We consider sinusoidal fluctuations along the $x$-axis
of the opacity $\kappa$ and Planck function $B_\nu$ of the form
\begin{align}
B(x, y, z) &= \bar{B}(z) + \Delta B(z) \cos kx, \label{eq21}
\\
\kappa(x, y, z) &= \bar{\kappa}(z)[1 + \alpha \cos kx].  \label{eq22}
\end{align}
If $\alpha >0$, the fluctuations are said to be in phase; 
if $\alpha <0$, the fluctuations are said to be in anti-phase.
For the linearization considerations of the previous section to apply, 
we assume $|\alpha|\ll 1$.

Due to the linearity of the problem the Stokes parameters have the following
spatial dependence\footnote{Higher harmonics $\cos(nkx)$ and $\sin(nkx)$, with $n\ge 2$, 
appear only if they are already present on the radiation field illuminating the boundaries. 
But then, lacking sources, they are exponentially attenuated 
and in a semi-infinite mean atmosphere, they vanish completely when approaching the surface.}
\begin{subequations}\label{eq23}
\begin{align}
\delta I(x, y, z) &= \Delta_1 I(z) \cos kx + \Delta_2 I(z) \sin kx, \label{eq23a}\\
\delta Q(x, y, z) &= \Delta_1 Q(z) \cos kx + \Delta_2 Q(z) \sin kx, \label{eq23b}\\
\delta U(x, y, z) &= \Delta_1 U(z) \cos kx + \Delta_2 U(z) \sin kx, \label{eq23c}
\end{align}
\end{subequations}
and 
\begin{subequations}\label{eq23_b}
\begin{align}
\delta S_I^{\rm eff}(x, y, z) &= [\Delta_1 S_I(z) \,-\,\alpha(\,\bar{I}-\bar{S}_I\,)]\cos kx + \Delta_2 S_I(z) \sin kx, \label{eq23_ba}\\
\delta S_Q^{\rm eff}(x, y, z) &= [\Delta_1 S_Q(z) -\alpha(\bar{Q}-\bar{S}_Q)]\cos kx + \Delta_2 S_Q(z) \sin kx, \label{eq23_bb}\\
\delta S_U^{\rm eff}(x, y, z) &= [\Delta_1 S_U(z) -\alpha(\bar{U}-\bar{S}_U)]\cos kx + \Delta_2 S_U(z) \sin kx. \label{eq23_bc}
\end{align}
\end{subequations}
The amplitudes $\Delta_{1, 2}S_I$, $\Delta_{1, 2}S_Q$, and $\Delta_{1, 2}S_U$, are expressed
in terms of $\Delta_{1, 2}S^0_0$, $\Delta_{1, 2}S^2_0$,..., $\Delta_{1, 2}\hat{S}^2_2$, as in Eqs.~(\ref{eq02}).
 

Now, we derive radiative transfer equations for the amplitudes $\Delta_{1, 2} I$,
$\Delta_{1, 2} Q$, and  $\Delta_{1, 2} U$ as follows.
In Cartesian geometry, taking the $y$-axis as the invariant direction, 
the explicit expression for the d/d$s$ operator in the radiative transfer equation is 
\begin{equation}
  \frac{\rm d}{{\rm d}s}\,\equiv\,\mu\frac{\partial}{\partial z}
	+\lambda\frac{\partial}{\partial x},
                                                              \label{eq24}
\end{equation}
with $\mu=\cos\theta$ and $\lambda=\sin\theta\cos\chi$ (see Fig.~\ref{fig01}).
From Eqs.~(\ref{eq17})-(\ref{eq18}) and Eqs.~(\ref{eq23})-(\ref{eq24})
we get, after some elementary algebra, the following coupled system of radiative 
transfer equations for the amplitudes:
\begin{align}
   \mu \frac{\rm d}{{\rm d} \tau} \Delta_1I &=\, \Delta_1I 
	-\big[\Delta_1S_I-\alpha(\bar{I}-\bar{S}_I)-\frac{\lambda k}{\bar{\kappa}}
	\Delta_2 I\big],
                                                              \label{eq25}
\displaybreak[0] \\
   \mu \frac{\rm d}{{\rm d} \tau} \Delta_2I &=\, \Delta_2I 
	-\big[\Delta_2S_I+\frac{\lambda k}{\bar{\kappa}}
	\Delta_1 I\big],
                                                              \label{eq26}
\end{align}
with analogous expressions for $\Delta_{1,2}Q$ and $\Delta_{1,2}U$.
A change of variable has been introduced in Eqs.~(\ref{eq25})-(\ref{eq26}), from the 
geometrical $z$-scale to the the monochromatic optical depth scale $\tau$, 
which is defined from the mean opacity
$\bar{\kappa}$, such that d$\tau=-\bar{\kappa}\,{\rm d}z$.
Note that unlike a {\em true} plane-parallel problem,
here, the radiation field depends explicitly on the azimuthal angle.
The apparent asymmetry between Eq.~(\ref{eq25}) and Eq.~(\ref{eq26}) (a term on $\alpha$ appears in the former but not on the latter), is just due to our choice of cosinusoidal opacity fluctuations in Eq.~(\ref{eq22}). 
If a more general dependence were chosen (e.g., $\cos(kx+\phi)$), an analogous term would then appear in Eq.~(\ref{eq26}) too.

The boundary conditions for the whole fluctuations
imply to the following boundary conditions for the amplitudes:
$\Delta_{1, 2}I_{\mu\chi}(z_M)=\Delta_{1, 2}Q_{\mu\chi}(z_M)
=\Delta_{1, 2}U_{\mu\chi}(z_M)=0$, for $\mu<0$, at the top; 
$\Delta_{1}I_{\mu\chi}(z_m) =\Delta B(z_m)$, $\Delta_{2}I_{\mu\chi}(z_m) =\Delta_{1, 2}Q_{\mu\chi}(z_m)
=\Delta_{1, 2}U_{\mu\chi}(z_m)=0$, for $\mu>0$, at the bottom.

\subsection{Statistical equilibrium equations}\label{sect42}

From Eqs.~(\ref{eq23}), the spatial dependence of the radiation field tensor perturbations is
\begin{equation}
\delta J^K_Q = \Delta_1 J^K_Q(z) \cos kx + \Delta_2 J^K_Q(z) \sin kx,  \label{eq31}
\end{equation}
where the amplitudes
$\Delta_{1, 2}J^K_Q(z)$ are computed from the 
corresponding $\Delta_{1, 2}I$, $\Delta_{1, 2}Q$, and $\Delta_{1, 2}U$ amplitudes 
according to Eqs.~(\ref{eq04}). 
Equation~(\ref{eq31}), in turn, implies the same spatial factorization for the $\delta S^K_Q$ fluctuations,
\begin{equation}
\delta S^K_Q = \Delta_1 S^K_Q(z) \cos kx + \Delta_2 S^K_Q(z) \sin kx,  \label{eq32}
\end{equation}
as can be seen from the statistical equilibrium equations of the problem under consideration.
Moreover, due to the linearity of the problem, the very same statistical equilibrium equations are 
satisfied, independently, for the terms fluctuating as $\cos kx$ ($\Delta_1 J^K_Q$, $\Delta_1 S^K_Q$), and the terms fluctuating as $\sin kx$ ($\Delta_2 J^K_Q$, $\Delta_2 S^K_Q$).

\begin{figure}[t]
\centerline{\epsfig{figure=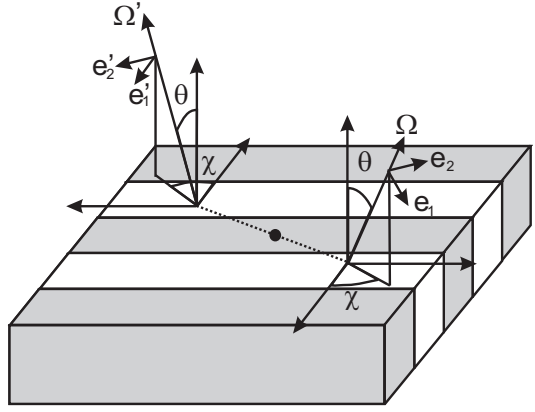, width=7cm}}
\caption{The atmosphere defined by Eqs.~(\ref{eq21})-(\ref{eq22})
is invariant under a 180$^\circ$ 
rotation around the origin of coordinates. 
Thus, radiation along a ray $\mvec{\Omega}$ at point $P(x, y, z)$, 
is identical to radiation along ray  $\mvec{\Omega'}$ at point $P(-x, -y, z)$. 
Therefore, $I_{\mu\chi}(x)=I_{\mu\chi+\pi}(-x)$, $Q_{\mu\chi}(x)=Q_{\mu\chi+\pi}(-x)$, 
and $U_{\mu\chi}(x)=U_{\mu\chi+\pi}(-x)$. 
\label{fig03}}
\end{figure}

\subsection{Further symmetry considerations}\label{sect43}

It is possible to exploit the additional symmetries arising in the problem from 
the particular fluctuation law we are considering (Eqs.~(\ref{eq21})-(\ref{eq22})).
From Fig.~\ref{fig03}, it is clear that 
$\delta I_{\mu\chi}(x)=\delta I_{\mu\chi+\pi}(-x)$, 
$\delta Q_{\mu\chi}(x)=\delta Q_{\mu\chi+\pi}(-x)$, and 
$\delta U_{\mu\chi}(x)=\delta U_{\mu\chi+\pi}(-x)$, which implies
\begin{align}
\Delta_1 I_{\mu\chi} &= \Delta_1 I_{\mu\chi+\pi}, \label{eq27}\\
\Delta_2 I_{\mu\chi} &= -\Delta_2 I_{\mu\chi+\pi},\label{eq28}
\end{align}
(the same is true for $\Delta_{1, 2} Q$ and $\Delta_{1, 2} U$).
Equations~(\ref{eq27})-(\ref{eq28}) lead to the following straightforward identities:
\begin{equation}
\int_0^{2\pi} {\rm d}\chi \; \Delta_2I_\chi =
\int_0^{2\pi} {\rm d}\chi \; \Delta_1I_\chi \sin\chi =
\int_0^{2\pi} {\rm d}\chi \; \Delta_2I_\chi \sin 2\chi =0,
\end{equation}
%
which apply to $\Delta_{1, 2} Q$ and $\Delta_{1, 2} U$ as well. 
When these relationships are taken into account
in the definitions of the radiation field tensors (Eqs.~(\ref{eq04})), 
we get 
\begin{subequations}\label{eq30}
\begin{align}
  \delta J^0_0&=\,\Delta_1 J^0_0(z)\cos(kx), 
                                                              \label{eq30a}
\displaybreak[0] \\
  \delta J^2_0&=\,\Delta_1 J^2_0(z)\cos(kx), 
                                                              \label{eq30b}
\displaybreak[0] \\
  \delta \tilde{J}^2_1&=\,\Delta_2 \tilde{J}^2_1(z)\sin(kx), 
                                                              \label{eq30c}
\displaybreak[0] \\
  \delta \tilde{J}^2_2&=\,\Delta_1 \tilde{J}^2_2(z)\cos(kx). 
                                                              \label{eq30d}
\end{align}
\end{subequations}
This factorization of the radiation field tensors 
is a consequence of the symmetry illustrated in Fig.~(\ref{fig03}).
This symmetry does not hold anymore in the presence of an inclined 
magnetic field; the radiation field tensors then fluctuate according to 
the more general Eqs.~(\ref{eq31}).
Note, in passing, that this would also be the case, even without magnetic fields, 
if the perturbed variables had a phase difference between them
(i.e., other than 0 or $\pi$), e.g., $\delta B \sim \cos kx$ and
$\delta \kappa \sim \sin kx$.

The most important point to keep in mind is that the general multi-dimensional problem has been reduced to a set
of plane-parallel coupled ones given by Eqs.~(\ref{eq25})-(\ref{eq26}) and their
corresponding set of statistical equilibrium equations. 
These {quasi-planeparallel} problems must now be solved numerically.

\section{Numerical method of solution}\label{sect5}

We iterate to obtain the self-consistent solution of the 
radiative transfer equations for the Stokes parameters amplitudes 
and the statistical equilibrium equations for the density matrix amplitudes.
Section~\ref{sect51} shows two different, but related, 
methods to integrate the RT equations to obtain the
Stokes parameters amplitudes and the radiation field tensor amplitudes 
$\Delta_{1, 2} J^K_Q$ when the $\Delta_{1, 2} S^K_Q$ elements are known.
Section~\ref{sect52} in turn, shows how to calculate new values of the 
$\Delta_{1, 2} S^K_Q$ amplitudes from the just computed radiation field amplitudes.
We explain how to iterate this process to reach convergence.

\subsection{Formal solution of the transfer equation}\label{sect51}

\subsubsection{Method (a)}\label{sect51a}

We integrate the RT equations along short characteristics (SC).
This method is based on the following strategy.
Let $M$, $O$, and $P$ be three consecutive points along a ray path.
Assuming that the source function varies parabolically between them, 
the intensity at point $O$ may be expressed as
\begin{equation}
I(O)=I(M){\rm e}^{-\Delta\tau_M}+\psi_M S_I(M)+\psi_O S_I(O)+\psi_P S_I(P), \label{eqsc}
\end{equation}
where $\Delta\tau_M$ is the optical distance between $M$ and $O$, and
$\psi_M$, $\psi_O$ and $\psi_P$ are coefficients that only depend on the 
optical distance between points $O$ and $M$ 
($\Delta\tau_M$), and $M$ and $P$ ($\Delta\tau_P$). 
The explicit functional form of the $\psi$ coefficients 
for linear or parabolic variations of the source function between grid points
can be found in Kunasz \& Auer (1988).
The linear version (without the $\psi_P$ term) must be used 
at the boundaries when no $P$ point is available.
Expressions analogous to Eq.~(\ref{eqsc}) are valid for the Stokes parameters $Q$ and $U$.
In planeparallel problems, $M$, $O$, and $P$ correspond to three consecutive
points on the $z$-grid. In two- and three-dimensional problems, $M$ and $P$ do
not correspond, in general, to grid points and some interpolation is needed 
(see Kunasz \& Auer 1988; Auer \& Paletou 1994; Auer, Fabiani Bendicho, \& Trujillo Bueno 1994;
Fabiani Bendicho \& Trujillo Bueno 1999).
Here, we avoid this computationally intensive interpolations by exploiting 
the known functional dependence of the variables on the horizontal direction.

Let $(x_L, y_L, z_L)$ be the Cartesian coordinates of point $L$. Noting that
$\Delta_1I(z_L)=\delta I(0, y_L, z_L)$, and $\Delta_2I(z_L)=\delta I(\frac{\pi}{2k}, y_L, z_L)$, 
then Eqs.~(\ref{eq23})-(\ref{eq23_b}), and Eq.~(\ref{eqsc}) imply\footnote{For the sake of clarity, only the case without opacity fluctuations ($\alpha=0$) is treated here explicitly.
The formal substitution $\Delta_1S_I\rightarrow \Delta_1S-\alpha(\bar{I}-\bar{S}_I)$ in these equations
yields the general expressions with opacity fluctuations.}
\begin{align}
  \begin{split}
     \Delta_1I(z_O)&=\,[\Delta_1I(z_M)\cos(\frac{\lambda k}{\mu}\Delta_M)-
       \Delta_2I(z_M)\sin(\frac{\lambda k}{\mu}\Delta_M)]{\rm e}^{-\Delta\tau_M}
     \\
     & \quad +\psi_M\big[\Delta_1S_I(z_M)\cos(\frac{\lambda k}{\mu}\Delta_M)
       -\Delta_2S_I(z_M)\sin(\frac{\lambda k}{\mu}\Delta_M)\big]+\psi_O\Delta_1S_I(z_O)
     \\
     & \quad +\psi_P\big[\Delta_1S_I(z_P)\cos(\frac{\lambda k}{\mu}\Delta_P)
       +\Delta_2S_I(z_P)\sin(\frac{\lambda k}{\mu}\Delta_P)\big],
  \end{split}
                                                              \label{eq33_}
\displaybreak[0] \\
  \begin{split}
     \Delta_2I(z_O)&=\,[\Delta_1I(z_M)\sin(\frac{\lambda k}{\mu}\Delta_M)+
       \Delta_2I(z_M)\cos(\frac{\lambda k}{\mu}\Delta_M)]{\rm e}^{-\Delta\tau_M}
     \\
     & \quad +\psi_M\big[\Delta_1S_I(z_M)\sin(\frac{\lambda k}{\mu}\Delta_M)
       +\Delta_2S_I(z_M)\cos(\frac{\lambda k}{\mu}\Delta_M)\big]+\psi_O\Delta_2S_I(z_O)
     \\
     & \quad +\psi_P\big[-\Delta_1S_I(z_P)\sin(\frac{\lambda k}{\mu}\Delta_P)
       +\Delta_2S_I(z_P)\cos(\frac{\lambda k}{\mu}\Delta_P)\big],
  \end{split}
                                                              \label{eq34_}
\end{align}
and analogously for $\Delta_{1, 2}Q$ and $\Delta_{1, 2}U$,
with $\Delta_M=z_O-z_M$ and $\Delta_P=z_P-z_O$. 
Note that the Stokes parameters amplitudes $\Delta_{1, 2}I$, $\Delta_{1, 2}Q$, and $\Delta_{1, 2}U$ depend on just one spatial dimension 
(depth), but they depend on both angular variables, $\mu=\cos\theta$ and $\lambda=\sin\theta\cos\chi$.
Proceeding from one boundary to the other along the ray direction, 
Eqs.~(\ref{eq33_})-(\ref{eq34_}) allow us to calculate
$\Delta_{1, 2}I$ (and $\Delta_{1, 2}Q$, and $\Delta_{1, 2}U$) at each height of the vertical grid, and from them the
Stokes parameters at any other horizontal position if necessary.

\subsubsection{Method (b)}\label{sect51b}

Equations (\ref{eq25})-(\ref{eq26}) are formally identical to the RT equation 
in a plane-parallel atmosphere if we identify the term between brackets on the 
right hand side with a source function.
We exploit this formal similarity to integrate numerically these equations 
through the well-known short-characteristics (SC) method for the formal solution of the RT equation.
Thus, integrating explicitly Eqs.~(\ref{eq25})-(\ref{eq26}) 
between three consecutive spatial points $M$, $O$ and $P$ along the ray
assuming a parabolic variation for the terms between brackets,
the intensity amplitudes at point $O$ can be expressed as\footnotemark[6]
\begin{align}
  \begin{split}
  \Delta_1I(z_O)&=\,\Delta_1I(z_M){\rm e}^{-\Delta\tau_M}
	+\psi_M\big[\Delta_1S_I(z_M)-\frac{\lambda k}{\bar{\kappa}(z_M)}\Delta_2I(z_M)\big] 
	\\ 
	& \quad +\psi_O\big[\Delta_1S_I(z_O)-\frac{\lambda k}{\bar{\kappa}(z_O)}\Delta_2I(z_O)\big]
	+\psi_P\big[\Delta_1S_I(z_P)-\frac{\lambda k}{\bar{\kappa}(z_P)}\Delta_2I(z_P)\big],
  \end{split}
                                                              \label{eq33}
\displaybreak[0] \\
  \begin{split}
  \Delta_2I(z_O)&=\,\Delta_2I(z_M){\rm e}^{-\Delta\tau_M}
	+\psi_M\big[\Delta_2S_I(z_M)+\frac{\lambda k}{\bar{\kappa}(z_M)}\Delta_1I(z_M)\big]
	\\ 
	& \quad +\psi_O\big[\Delta_2S_I(z_O)+\frac{\lambda k}{\bar{\kappa}(z_O)}\Delta_1I(z_O)\big]
	+\psi_P\big[\Delta_2S_I(z_P)+\frac{\lambda k}{\bar{\kappa}(z_P)}\Delta_1I(z_P)\big],
  \end{split}
                                                              \label{eq34}
\end{align}
with similar expressions for $\Delta_{1, 2}Q$ and $\Delta_{1, 2}U$.
Note that here, as in Eqs.~(\ref{eq33_})-(\ref{eq34_}), $\Delta\tau_M$ 
is calculated from the mean opacity $\bar{\kappa}$. 

{In contrast with} a true plane-parallel problem, Eqs.~(\ref{eq33})-(\ref{eq34}) 
do not give explicitly the values of $\Delta_1I$ and $\Delta_2I$ at a given point $O$  
once the intensity at the previous point $M$ is known. Instead,  
$\Delta_1I$ and $\Delta_2I$ at $O$ depend on those very same quantities at points $M$ 
{\em and} $P$.
The intensity amplitudes are given implicitly, in the form of an algebraic 
system of equations that can be written in matrix form as
\begin{equation}
  -\mvec{A}_O\mvec{v}_M+\mvec{B}_O\mvec{v}_O-\mvec{C}_O\mvec{v}_P\,=\,
	\mvec{b}_O,
                                                              \label{4.lin.30}
\end{equation}
where
\begin{gather}
   \mvec{A}_O \,=\, \left( \begin{array}{cc}
	{\rm e}^{-\Delta\tau_M} & -\psi_M \frac{\lambda k}{\bar{\kappa}(z_M)} \\
	\psi_M \frac{\lambda k}{\bar{\kappa}(z_M)} & {\rm e}^{-\Delta\tau_M} 
	\end{array} \right),				      	
   \mvec{B}_O \,=\, \left( \begin{array}{cc}
	1 & \psi_O \frac{\lambda k}{\bar{\kappa}(z_O)} \\
	-\psi_O \frac{\lambda k}{\bar{\kappa}(z_O)} & 1 
	\end{array} \right), \nonumber \displaybreak[0] \\
   \mvec{C}_O \,=\, \left( \begin{array}{cc}
	0 & -\psi_P \frac{\lambda k}{\bar{\kappa}(z_P)} \\
	\psi_P \frac{\lambda k}{\bar{\kappa}(z_P)} & 0 
	\end{array} \right), \nonumber \displaybreak[0] \\
   \mvec{b}_O \,=\, \left( \begin{array}{c}
	\psi_M\Delta_1S_I(z_M)+\psi_O\Delta_1S_I(z_O)+\psi_P\Delta_1S_I(z_P) \\
	\psi_M\Delta_2S_I(z_M)+\psi_O\Delta_2S_I(z_O)+\psi_P\Delta_2S_I(z_P)
	\end{array} \right), 		\nonumber 	       
\end{gather}
and $\mvec{v}_M$, $\mvec{v}_O$ and $\mvec{v}_P$ are vectors whose two 
components are the amplitudes
$\Delta_1I$ and $\Delta_2I$ at points $M$, $O$ and $P$, respectively.
By changing the corresponding values in vectors $\mvec{b}$ and $\mvec{v}$,
the expressions apply to all  Stokes parameters.
The set of Eqs.~(\ref{4.lin.30}) for all points in the atmosphere
form a 2$\times$2-blocks tridiagonal system of equations.
Such a system can be efficiently solved through a standard 
{\em forth and back} substitution scheme (e.g., Press \etal 1992):
\begin{gather}
  \mvec{v}_i\,=\,\mvec{D}_i\mvec{v}_{i+1}+\mvec{u}_i, 
                                                              \label{4.lin.31}
\displaybreak[0] \\
  \mvec{D}_i\,\equiv\,(\mvec{B}_i-\mvec{A}_i\mvec{D}_{i-1})^{-1}\mvec{C}_i,
                                                              \label{4.lin.32}
\displaybreak[0] \\
  \mvec{u}_i\,\equiv\,(\mvec{B}_i-\mvec{A}_i\mvec{D}_{i-1})^{-1}
	(\mvec{b}_i+\mvec{A}_i\mvec{u}_{i-1}),
                                                              \label{4.lin.33}
\end{gather}
where $i=1, ..., N_z$ ($N_z$ is the number of spatial grid points). We start at one atmosphere's boundary ($i=1$)
and calculate $\mvec{D}_i$ and $\mvec{u}_i$ at all grid points
until $i=N_z-1$. At the other boundary $i=N_z$, 
$\mvec{C}_{N_z}=\mvec{D}_{N_z}=0$
and $\mvec{v}_{N_z}=\mvec{u}_{N_z}$. Then we proceed backwards 
obtaining $\mvec{v}_i$ from equation (\ref{4.lin.31}).
This process involves the inversion of 2$\times$2 matrices
which can be performed analytically, and only requires a little extra
numerical effort.

Both methods of integration of the RT equation presented in this section
converge to the same solution for increasingly finer grids (i.e., as $\Delta z\rightarrow 0$).
Method (a) assumes a parabolic variation of the source functions perturbations
$\delta S_i(x, y, z)$ between three consecutive points, 
regardless of the known factorization given in Eqs.~(\ref{eq23}) .
Method (b) assumes a parabolic variation of the source functions amplitudes
$\Delta_{1, 2} S_i(z)$ between three consecutive heights, 
and the horizontal variation of the physical parameters is taken into account
explicitly on the derivation of the equations. 
Therefore, method (a) is easier to implement and slightly faster (a tridiagonal system
of equations is avoided), while method (b) is more accurate for coarser spatial grids.
The difference between both is more evident for higher $k$, 
and very inclined rays close to the $y$-$z$ plane. 
In that case, rays may cross several horizontal periods of the fluctuating
source functions, which is poorly approximated by a parabolic variation.
Therefore, method (a) requires finer $z$-grids at high wavenumbers $k$.

\subsection{Iterative scheme}\label{sect52}
 
Equations (\ref{eq33_})-(\ref{eq34_}) or Eqs.~(\ref{eq33})-(\ref{eq34}) show a linear 
relationship between the source functions and Stokes parameters amplitudes. 
Formally:
\begin{subequations}\label{eq50}
\begin{align}
  \mvec{\delta I}&=\,\mvec{\Lambda}_{\mvec{\Omega}}
	[\mvec{\delta S}_I]+\mvec{\delta {\cal I}}, 
                                                              \label{eq50a}
\displaybreak[0] \\
  \mvec{\delta Q}&=\,\mvec{\Lambda}_{\mvec{\Omega}}
	[\mvec{\delta S}_Q]+\mvec{\delta {\cal Q}}, 
                                                              \label{eq51b}
\displaybreak[0] \\
  \mvec{\delta U}&=\,\mvec{\Lambda}_{\mvec{\Omega}}
	[\mvec{\delta S}_U]+\mvec{\delta {\cal U}},
                                                              \label{eq52b}
\end{align}
\end{subequations}
where $\mvec{\delta I}$, $\mvec{\delta Q}$, and $\mvec{\delta U}$ are
2$\times N_z$-element vectors whose components are the $\Delta_{1, 2} I$,
$\Delta_{1, 2} Q$, $\Delta_{1, 2} U$ amplitudes at the $N_z$ heights in the
atmosphere, respectively. 
The 2$\times N_z$ components of $\mvec{\delta {\cal I}}$, $\mvec{\delta {\cal Q}}$, 
and $\mvec{\delta {\cal U}}$ are the corresponding Stokes parameters amplitudes
transmitted from the boundary.
$\mvec{\delta S}_I$, $\mvec{\delta S}_Q$, and
$\mvec{\delta S}_U$ are 2$\times N_z$-element vectors with components
$\Delta_{1, 2} S_I$, $\Delta_{1, 2} S_Q$, and $\Delta_{1, 2} S_U$, respectively.
Finally, $\mvec{\Lambda}_{\mvec{\Omega}}$ is a
$(2\times N_z)\times(2\times N_z)$-element matrix;
the same for all three Stokes parameters that depends on both,
the inclination (through $\mu$) and azimuthal angle (through $\lambda$)
of the ray.
The ${\Lambda}_{\mvec{\Omega}}(i, j)$ elements are involved functions
of the optical distances between spatial grid points that
need not be evaluated explicitly ---they are implicitly calculated 
every time a formal solution is performed as in the previous section.
However, Eqs.~(\ref{eq50}) show that they can be numerically evaluated 
straightforwardly.
If the medium is not illuminated at the boundaries 
($\mvec{\delta {\cal I}}=\mvec{\delta {\cal Q}}=\mvec{\delta {\cal U}}=0$),
and we consider a source function perturbation vanishing at all points
but at point $i$, where $\delta S_I(i)=1$ (or $\delta S_Q(i)=1$ or $\delta S_U(i)=1$),
then the corresponding Stokes parameter fluctuation calculated at 
point $j$ as in Sect.~\ref{sect51} is $\delta I(j)=\Lambda_{\mvec{\Omega}}(i, j)$
($\delta Q(j)=\Lambda_{\mvec{\Omega}}(i, j)$, $\delta U(j)=\Lambda_{\mvec{\Omega}}(i, j)$).
In particular, the diagonal element ${\Lambda}_{x\mvec{\Omega}}(i, i)$ is
the intensity amplitude at a given point when the corresponding source function
amplitude at that point is 1 and vanish elsewhere. 

We introduce the 2$\times N_z$-element vectors $\mvec{\delta S^0_0}$, 
$\mvec{\delta S^2_0}$, ..., $\mvec{\delta \hat{S}^2_2}$, and
$\mvec{\delta J^0_0}$, $\mvec{\delta J^2_0}$, ..., $\mvec{\delta \hat{J}^2_2}$,
whose components are 
$\Delta_{1, 2} S^0_0$, ..., $\Delta_{1, 2} \hat{S}^2_2$, 
$\Delta_{1, 2} J^0_0$, ..., $\Delta_{1, 2} \hat{J}^2_2$, respectively, 
at each point in the atmosphere.
The source functions depend linearly on the $S^K_Q$ components (Eqs.~(\ref{eq02})),
while the $J^K_Q$ elements are weighted angular averages of the Stokes parameters
(Eqs.~\ref{eq04}). 
Formally:
\begin{subequations}\label{eq53}
\begin{align}\label{eq53a}
\mvec{\delta J^0_0}&=\mvec{\Lambda}_{00}\mvec{\delta S^0_0}+
\mvec{\Lambda}_{01}\mvec{\delta S^2_0}+
\mvec{\Lambda}_{02}\mvec{\delta \tilde{S}^2_1}+
\mvec{\Lambda}_{04}\mvec{\delta \tilde{S}^2_2}+
\mvec{\Lambda}_{0c}\mvec{\delta S^{cont}}+
\mvec{\delta {\cal J}^0_0}, 
\\
\mvec{\delta J^2_0}&=\mvec{\Lambda}_{10}\mvec{\delta S^0_0}+
\mvec{\Lambda}_{11}\mvec{\delta S^2_0}+
\mvec{\Lambda}_{12}\mvec{\delta \tilde{S}^2_1}+
\mvec{\Lambda}_{14}\mvec{\delta \tilde{S}^2_2}+
\mvec{\Lambda}_{1c}\mvec{\delta S^{cont}}+
\mvec{\delta {\cal J}^2_0},                                           \label{eq53b}
\\
\mvec{\delta \tilde{J}^2_1}&=\mvec{\Lambda}_{20}\mvec{\delta S^0_0}+
\mvec{\Lambda}_{21}\mvec{\delta S^2_0}+
\mvec{\Lambda}_{22}\mvec{\delta \tilde{S}^2_1}+
\mvec{\Lambda}_{24}\mvec{\delta \tilde{S}^2_2}+
\mvec{\delta \tilde{{\cal J}}^2_1},                                           \label{eq53c} 
\\
\mvec{\delta \hat{J}^2_1}&=
\mvec{\Lambda}_{33}\mvec{\delta \hat{S}^2_1}+
\mvec{\Lambda}_{35}\mvec{\delta \hat{S}^2_2}+
\mvec{\delta \hat{{\cal J}}^2_1},                                            \label{eq53d}
\\
\mvec{\delta \tilde{J}^2_2}&=\mvec{\Lambda}_{40}\mvec{\delta S^0_0}+
\mvec{\Lambda}_{41}\mvec{\delta S^2_0}+
\mvec{\Lambda}_{42}\mvec{\delta \tilde{S}^2_1}+
\mvec{\Lambda}_{44}\mvec{\delta \tilde{S}^2_2}+
\mvec{\delta \tilde{{\cal J}}^2_2},                                            \label{eq53e}
\\
\mvec{\delta \hat{J}^2_2}&=
\mvec{\Lambda}_{53}\mvec{\delta \hat{S}^2_1}+
\mvec{\Lambda}_{55}\mvec{\delta \hat{S}^2_2}+
\mvec{\delta \hat{{\cal J}}^2_2},                                           \label{eq53f}
\end{align}
\end{subequations}
where the matrices $\mvec{\Lambda}_{\alpha\beta}$ 
are obtained from $\mvec{\Lambda}_{\mvec{\Omega}}$:
\begin{align}
   \Lambda_{\alpha\beta}(i, j) &=\, \int {\rm d}\nu \,\phi_\nu(i) r_\nu(j)
	\oint \frac{{\rm d}\vec{\Omega}}{4\pi} 
	\varpi^{\alpha\beta}_\vec{\Omega} \Lambda_{\vec{\Omega}}(i, j),\label{eq54}
\\ 
   \Lambda_{\alpha c}(i, j) & =\, \int {\rm d}\nu \,\phi_\nu(i) [1-r_\nu(j)]
	\oint \frac{{\rm d}\vec{\Omega}}{4\pi} 
	\varpi^{\alpha c}_\vec{\Omega} \Lambda_{\vec{\Omega}}(i, j).\label{eq55}
\end{align}
The angular factors $\varpi^{\alpha\beta}_\vec{\Omega}$ and 
$\varpi^{\alpha c}_\vec{\Omega}$ are obtained after some simple algebraic 
manipulations (see Appendix) and are given in Tables~\ref{tab4.3} and
\ref{tab4.4}, respectively. 
Note that due to symmetry arguments raised previously, 
Eqs.~(\ref{eq53d}) and (\ref{eq53f}) are relevant only if a magnetic
field is present.

Let $\mvec{\delta S^K_Q}^{\rm old}$ be the value of $\mvec{\delta S^K_Q}$ 
at some previous ``old'' iterative step and let $\mvec{\delta J^K_Q}^{\rm old}$ be 
the vector of radiation field tensor amplitudes calculated from 
$\mvec{\delta S^K_Q}^{\rm old}$ by formal integration of the RT equation (Sect.~\ref{sect51}). 
We seek to calculate the ``new'' values $\mvec{\delta S^K_Q}^{\rm new}$ of the density matrix amplitudes 
at the following iterative step.
We proceed by a variation of the approximate $\Lambda$-iteration 
(ALI) method along the lines used by Manso Sainz \& Trujillo Bueno (1999), and 
Manso Sainz (2002). The new amplitude value at a given point ``$i$'' in the 
atmosphere is obtained by solving the statistical equilibrium equations 
of the problem (Eqs.~(\ref{eq03}) or (\ref{eq10}))
with the following estimates for the radiation field tensors:
\begin{align}
   \delta J^0_0(i) & \approx \delta J^0_0(i)^{\rm old} +
	\Lambda_{00}(i, i)[\delta S^0_0(i)^{\rm new}-\delta S^0_0(i)^{\rm old}], \label{eq56}
\\ 
   \delta J^2_Q(i) & \approx \delta J^2_Q(i)^{\rm old}.                      \label{eq57}
\end{align}
Substituting Eq.~(\ref{eq56}) into Eq.~(\ref{eq03a}) (or (\ref{eq10a})) leads to
\begin{equation}\label{eq58}
\delta S^0_0(i)^{\rm new} = \frac{(1-\epsilon)[\delta J^0_0(i)^{\rm old}
-\Lambda_{00}(i, i)\delta S^0_0(i)^{\rm old}]
+\epsilon B_\nu}{1-(1-\epsilon)\Lambda_{00}(i, i)},
\end{equation}
which is the well-known ALI correction (see Trujillo Bueno \& Manso Sainz 1999, and references therein).
Substituting Eq.~(\ref{eq57}) into Eqs.~(\ref{eq03b})-(\ref{eq03f}) or 
Eqs.~(\ref{eq10b}))-(\ref{eq10f}) leads to a $\Lambda$-iteration correction
for $\delta S^2_0$, $\delta\tilde{S}^2_1$, ..., $\hat{S}^2_2$.

This is the simplest convergent iterative scheme possible for this problem.
The iterative scheme can be generalized by extending the procedure on 
$\Lambda_{00}$ in Eq.~(\ref{eq56}) to any other operator $\Lambda_{\alpha\beta}$ 
in Eq.~\ref{eq53}.
This leads to slightly more complex expressions of the type of Eq.~(\ref{eq58}) 
for all $\delta S^K_Q$ elements. However, 
there is little gain in the convergence rate 
for the particular problems we considered here.

\begin{figure}\label{fig03.5}
\centerline{\epsfig{figure=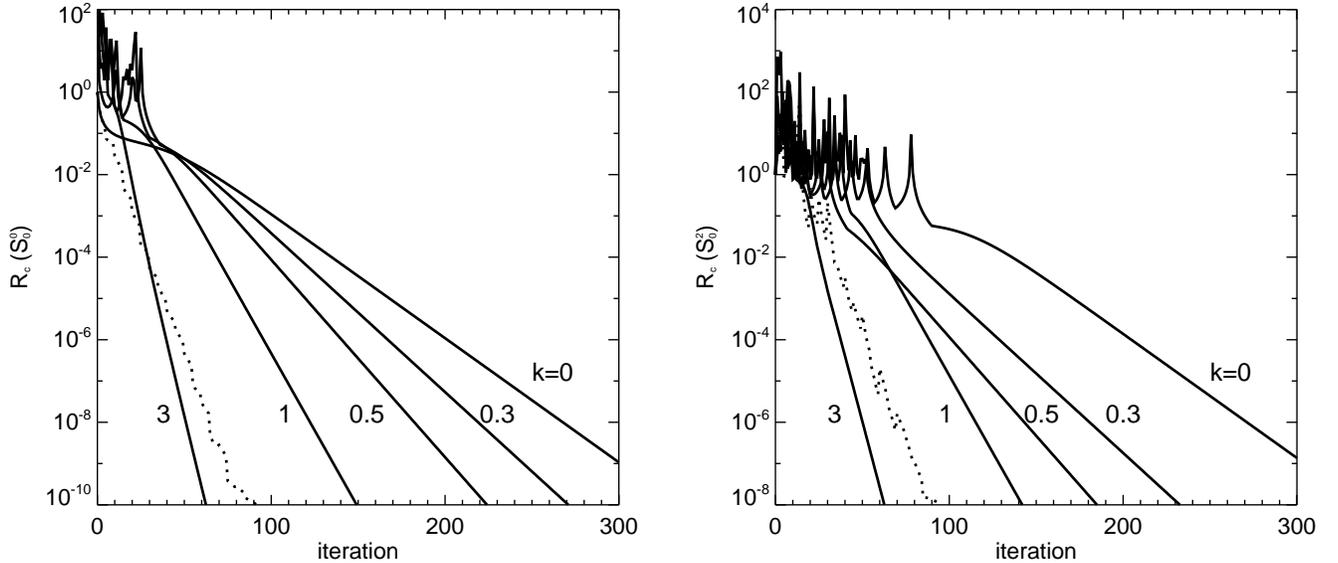, width=18cm}}
\caption{Maximum relative change $R_C(S^0_0)$ (left panel) and $R_C(S^2_0)$ (right panel), 
for different values of the horizontal wavenumber $k$ (labels). 
Scattering polarization in a non magnetized atmosphere with constant properties (and identical
for all cases) is considered. Dotted line: convergence rate for $k=0$ with Ng-acceleration.}
\end{figure}

Figure~\ref{fig03.5} shows the maximum relative change 
\begin{equation}
R_C=\max_i |\frac{\delta S^K_Q(i)^{\rm new}-\delta S^K_Q(i)^{\rm old}}{\delta S^K_Q(i)^{\rm new}}|
\end{equation}
of $S^0_0$ and $S^2_0$ in a scattering polarization problem. 
Different horizontal wavenumbers are considered. It can be shown that the convergence 
rate improves for larger horizontal wavenumbers. This is because effectively, the medium becomes more and
more optically thin. 
It can also be seen that a factor $\sim$3 improvement is obtained by just performing Ng-acceleration
to the standard ALI iterative scheme.
Comparing both panels we see that the convergence rate of the $S^2_0$ elements is tied to 
the convergence of the populations ($S^0_0$).

The convergence rate is independent of the method (a or b) employed for the formal solution integration.

\subsection{Numerical Considerations}\label{sect53}

\begin{figure}\label{fig04.5}
\centerline{\epsfig{figure=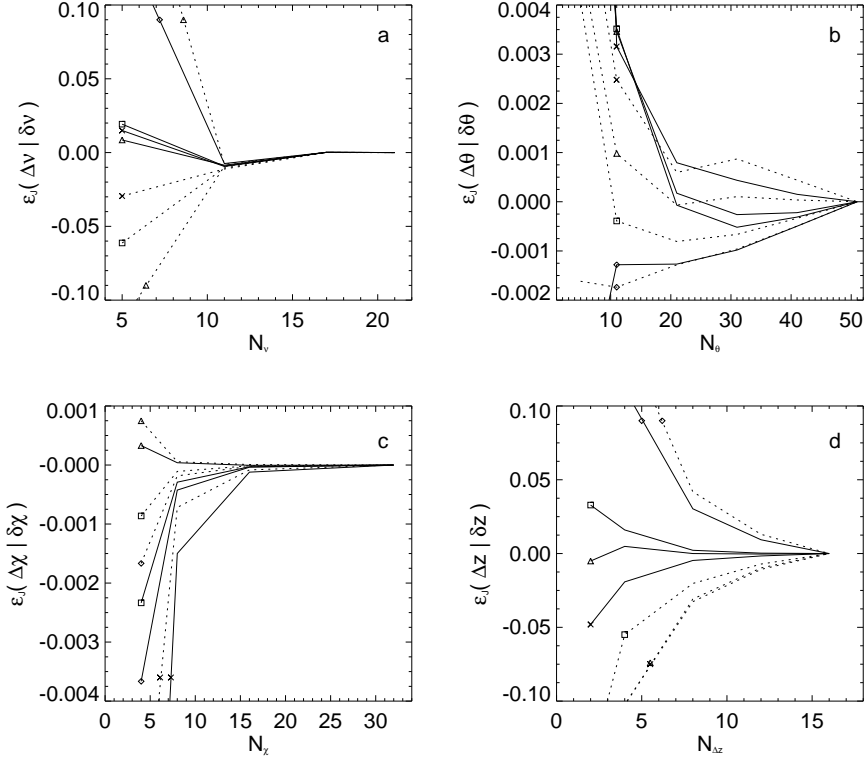, width=12cm}}
\caption{Relative discretization error for the value of $\Delta J^0_0/\bar{B}_\nu$ ($\Diamond$), 
$\Delta J^2_0/\bar{J}^0_0$ ($\triangle$), $\Delta J^2_1/\bar{J}^0_0$ ($\square$), and $\Delta J^2_2/\bar{J}^0_0$ ($\times$), at $z=0$ (solid lines), and $z=2$ (dotted lines), for an atmosphere with horizontal fluctuations with $k=0.3$ of the source function. 
The relative errors are given with respect to the solution given in Table~3. 
Panel a: frequency discretization, $N_\nu$ is the number of frequency points between the core and 10 $\Delta\nu_D$ Panel b: polar angle discretization, $N_\theta$ is the number of Gaussian points Panel c: azimuthal angle discretization, $N_\chi$ is the number of points for a trapezoidal rule between 0 and $2\pi$. Panel d: spatial discretization, $N_\Delta z$ is the number of grid points between $z$ and $z+1$.
}
\end{figure}

\subsubsection{Angular quadrature and symmetry}

The angular quadrature in multidimensional radiative transfer problems requires 
special consideration. This is even more so when dealing with polarization.

Kneer and Heasley (1979) introduced an efficient numerical quadrature
to deal with the complicated angular dependencies of the radiation field
in the unpolarized limit of the problems considered here (sinusoidal fluctuations of opacity and Planck function).
It consists of a Gaussian quadrature for polar angles taken from the invariant direction 
(here, the $y$-axis), and a trapezoidal rule for the azimuth on the orthogonal plane. 
Unfortunately, this quadrature is no well suited for polarization purposes.
In order to capture the symmetries of the problem and to be able to recover the plane-parallel limit 
(both, when $k=0$ and $k\rightarrow\infty$), the azimuthal quadrature must satisfy
the following relations {\em exactly}:
\begin{equation}\label{asdf}
\Sigma_i\cos\chi_i = \Sigma_i\sin\chi_i =\Sigma_i\cos2\chi_i = \Sigma_i\sin2\chi_i = 0,
\end{equation}
where the sum extends over all the directions with a fixed inclination with respect to
the vertical.
When Eqs.~(\ref{asdf}) are not satisfied exactly, spurious sources of polarization
appear.

The simplest quadrature satisfying these properties is a trapezoidal 
rule with at least four points uniformly distributed between 0 and $2\pi$. 
We use a $N_\mu$-point Gaussian quadrature for the inclination and a $N_\chi$-point trapezoidal rule for the azimuth.
As pointed out by Kneer and Heasley (1979), this choice has the inconvenience of 
not resolving adequately the variations of the radiation field due to 
sinusoidal fluctuations of the source function. 
Consequently, relatively fine grids are needed at moderate $k$ to avoid 
oscillations very close to the surface, where the medium becomes optically thin.
This inconvenience is {overcome} by our not introducing spurious sources of polarization.

The precision of the finite grid discretization is quantified by 
\begin{equation}
\epsilon_F(\Delta v|\delta v)=\frac{F_{[\Delta v]}-F_{[\delta v]}}{F_{[\delta v]}}, 
\end{equation}
which gives the value of some quantity $F$ in a coarse grid of a given variable ($\Delta v$)
relative to its value in a finer grid ($\delta v$), all other quantities kept equal.
Panels b and d of Fig.~\ref{fig04.5} show the $\epsilon_F(\Delta\theta|\delta\theta)$
and $\epsilon_F(\Delta\chi|\delta\chi)$ for $F=\Delta J^0_0/\bar{B}_\nu$, $\Delta J^2_0/\bar{J}^0_0$, $\Delta J^2_1/\bar{J}^0_0$, and $\Delta J^2_2/\bar{J}^0_0$, at $z=0$ and $z=2$.
The reference grid is the one for the results given in Fig.~8 and Table~3.

\subsubsection{Frequency quadrature}

We consider Gaussian line profiles $\phi_\nu$ and a static medium. 
A trapezoidal rule is used for the frequency integration.
Panel a in  Fig.~\ref{fig04.5} show $\epsilon_F(\Delta\nu|\delta\nu)$ as a function
of the number of points $N_\nu$ between $\nu_1=0$ (line core), and $\nu_{N_\nu}=10\Delta\nu_D$ ($\Delta\nu_D$ being the Doppler width).

\subsubsection{Spatial (vertical) discretization}

In plane-parallel media only the optical depth is a relevant variable,
the actual distribution of opacity with height is unimportant; 
in two and three dimensions however, 
the stratification of opacity with geometrical depth becomes relevant too.
The atmosphere's gravitational stratification suggests an exponential mean opacity 
$\bar{\kappa}(z)={\rm e}^{-z}$. Here, $\bar{\kappa}$ is normalized 
to its value at $z=0$ and the spatial variables $z$ and $x$ are
given in units of the opacity scale height ${\cal H}$. 
For reference, in the solar photosphere ${\cal H}$ is between $\approx 100$ and $200$~km
for typical spectral lines.  
The (vertical) optical depth (see Sect.~\ref{sect41}) is then $\tau={\rm e}^{-z}-{\rm e}^{-z_{\rm max}}$.

We consider a spatial grid $\{z_k \}_{k=1, ..., N_z}$ between $z_{\rm min}=-12$ and $z_{\rm max}=18$.
Therefore, $\tau\approx {\rm e}^{-z}$ in units of ${\cal H}$; hence,
$\tau_{\rm min}=10^{-8}$ and $\tau_{\rm max}=1.6\times 10^5$.
Panel c in  Fig.~\ref{fig04.5} show the relative error $\epsilon_F(\Delta z|\delta z)$
with respect to the grid necessary to obtain the results in  Fig.~8 and Table~3 as a function of the number of $z$ grid
points per unit $z$ interval.

\subsubsection{Accuracy and performance}

When we consider horizontal fluctuations only on the Planck function the
radiative transfer problem posed is already linear and the methods of solution 
presented in this section (in particular, the formal solution) are {\em exact}.
Furthermore, since no horizontal discretization is needed, the solution
obtained is more accurate than the solution obtained with other methods 
(based either on finite differences or SC) 
that require the discretization of all three spatial coordinates 
(e.g., Auer, Fabiani Bendicho, \& Trujillo Bueno 1994; Manso Sainz \& Trujillo Bueno 1999).
Thus, for example, such `classical' methods approach the solutions
of the problems considered in Sects. 6.1 and 6.2 below 
only in the limit of very fine horizontal grids. 

Treating the horizontal fluctuations exactly has the additional advantage of
improving the numerical performance by avoiding costly interpolations.
SC methods require knowledge of the source function at points $M$ and $P$,
and the intensity at point $M$ (see Eq.~(\ref{eqsc}). 
But for very special configurations, in two- and three-dimensional spatial grids, 
the $M$ and $P$ points do not belong to the spatial grid and the 
relevant quantities must be interpolated from neighboring grid point values.  
This requires that the interpolation coefficients be calculated 
for every direction of the angular quadrature. 
This calculation scales at best\footnote{For very inclined rays, or when the horizontal grid is relatively fine compared to the vertical grid, long-characteristics integrations might be necessary, worsening the problem of interpolation by so doing across several different cells for every single ray (e.g., Auer et al. 1994).}  
as $\sim N_\Omega\times N_P$ ($N_\Omega=N_\mu \times N_\chi$ being
the total number of directions of the angular quadrature; $N_P=N_x$, or $N_x\times N_y$ in two and three dimensions, respectively, $N_x$ and $N_y$ being the total number of points along the respective axes),
and should be repeated at each iterative step (everytime a formal solution is performed),
or otherwise, stored in the computer's memory.

When the opacity fluctuates horizontally, the radiative transfer problem is linearized
along the lines of Sect.~3. 
The numerical considerations just discussed apply to the linearized problem too:
accuracy and performance improve due to the lack of horizontal discretization.
However, there is an intrinsic loss of precission in the solution from the linearization 
of RT equation. The relative error in the solution is of the order of the second order 
terms neglected in the linearization: $\alpha\delta I$ for the amplitude $\delta I$ 
(and analogously for the other Stokes parameters amplitudes).
Typical errors in the amplitudes of the profiles for the problems considered
in Sects.~6.3 and 6.4 are below 1-0.1\%.

\section{Results and Discussion}\label{sect6}

We study a few selected, idealized problems: 
coherent continuum scattering polarization in the presence of horizontal fluctuations of the temperature (Sect. \ref{sect62}),
resonance scattering polarization with horizontal fluctuations of the temperature (Sect. \ref{sect61}), 
resonance scattering polarization with horizontal fluctuations of the opacity and source function (Sect. \ref{sect63}), 
and Hanle effect with horizontal fluctuations of the opacity and source function (Sect. \ref{sect64}).  
In particular, we consider cosinusoidal fluctuations of the parameters and we study the behavior of the solutions and emergent Stokes polarization profiles with the wavenumber.

\subsection{Continuum scattering polarization, source function fluctuations}\label{sect62}

We consider coherent scattering polarization in a constant properties, semi-infinite atmosphere with sinusoidal, horizontal fluctuations of the Planck function.

The total opacity (absorption plus scattering) is exponentially stratified with height. 
We normalize spatial distances to the opacity scale height such that $\kappa=\kappa^{\rm cont}+\sigma=\kappa_\circ{\rm e}^{-z}$, with $\kappa^{\rm cont}/\sigma$ constant throughout the atmosphere.
An arbitrary height reference is set by choosing $\kappa_\circ=1$.
The optical depth from the surface ($z\rightarrow\infty$), at a given $\mu$ is then $\tau={\rm e}^{-z}/\mu$; conversely, the height (in opacity scale units) at which $\tau=1$ along a line-of-sight (LOS) with $\mu$ is $z=\log(1/\mu\sqrt{\pi})$. 
Finally, we normalize $\Delta B$, $\Delta J^0_0$, $\Delta S^0_0$ amplitudes to the mean Planck function.

The solution of this problem for $\kappa^{\rm cont}/(\kappa^{\rm cont}+\sigma)=10^{-4}$ and a Planck function fluctuation 
$\Delta B/\bar{B}=1$ is given in Fig.~\ref{fig-05} and Table~\ref{tab2}.
They show the variation of $\Delta J^0_0$, $\Delta J^2_0$, $\Delta\tilde{J}^2_1$, and $\Delta\tilde{J}^2_2$ 
with height as a function of the horizontal wavenumber $k$.
Due to linearity, the solution for any value of $\Delta B/\bar{B}$ is obtained as the linear combination of the plane-parallel case (here, the $k=0$ solution) and $\Delta B/\bar{B}$ times this one.

\begin{figure}
\centerline{\epsfig{figure=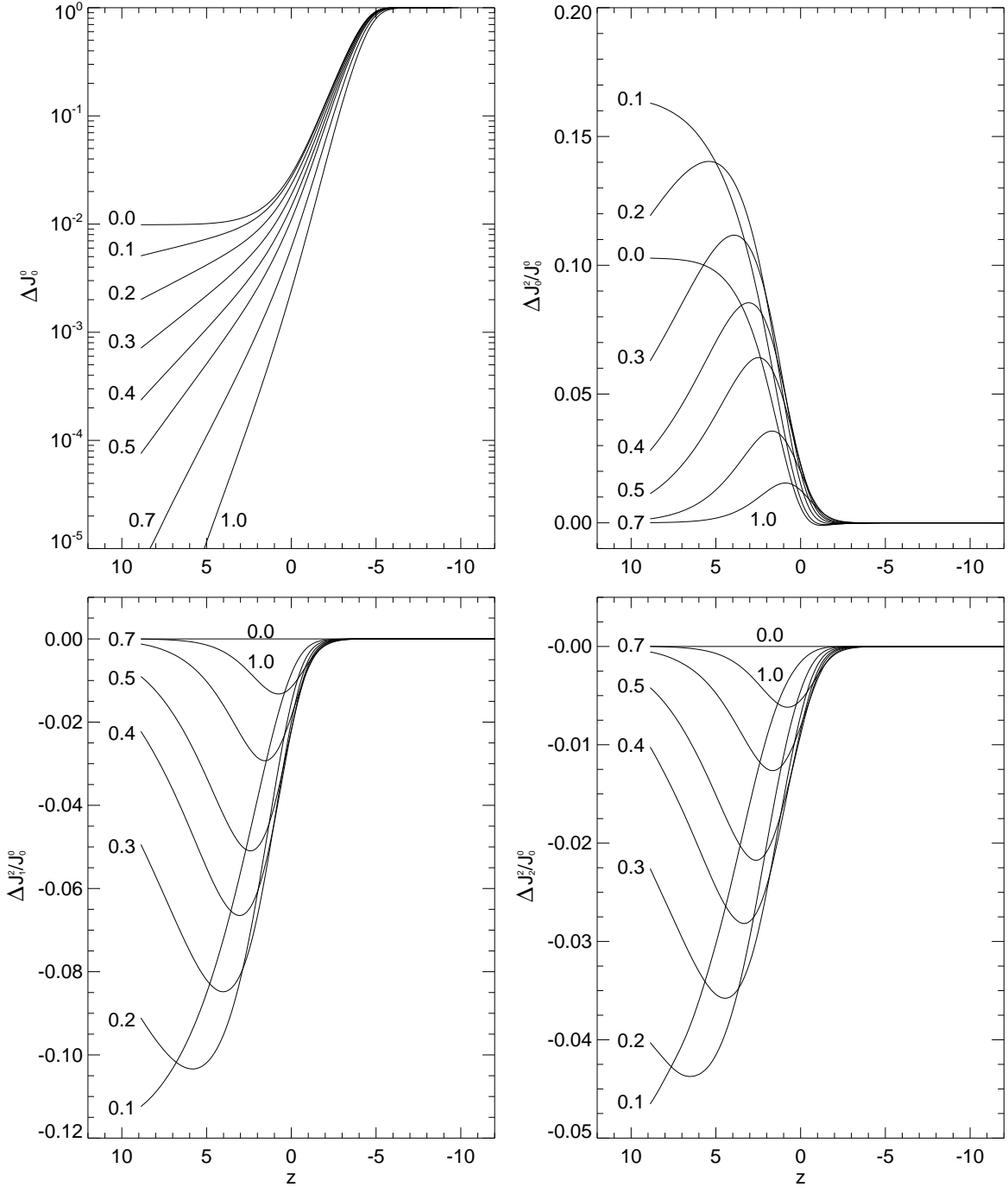, width=16cm}}
\caption{Polarization by coherent scattering in an atmosphere with horizontal, cosinusoidal 
fluctuations (labels indicate wavenumber $k$) of the Planck function (Eq.~(\ref{eq21})). 
An exponentially stratified atmosphere with $\kappa^{\rm cont}/(\sigma+\kappa^{\rm cont})=10^{-4}$, 
$\Delta B/\bar{B}=1$ is considered.
From top to bottom, and from left to right: $\Delta J^0_0$, $\Delta J^2_0/\bar{J}^0_0$,
$\Delta \tilde{J}^2_1/\bar{J}^0_0$, $\Delta \tilde{J}^2_2/\bar{J}^0_0$, respectively, as a function of 
height ($z$) measured in units of the opacity scale height ($z=0$ corresponds to $\tau=1$). \label{fig-05}}
\end{figure}

\begin{table}
\caption{Radiation field tensor amplitudes for $\kappa^{\rm cont}/(\kappa^{\rm cont}+\sigma)=10^{-4}$\label{tab2}}
\begin{tabular}{r.......}
\hline\\[-4pt]
\multicolumn{8}{c}{$\Delta J^0_0/B$} \\[4pt]
\hline\\[-4pt]
 & & \multicolumn{6}{c}{$k$} \\
z & \multicolumn{1}{c}{$\tau$} & 0.1 & 0.2& 0.3 & 0.5 & 0.7 & 1.0 \\
\hline\\[-4pt]
  8 & 0.00033   &  5.50(-3) &  2.36(-3) &  9.1(-4) &  1.1(-4) &  1.3(-5) &   \\
  6 & 0.0025    &  6.59(-3) &  3.44(-3) &  1.62(-3) &  3.1(-4) &  5.3(-5) &   \\
  4 & 0.018     &  8.12(-3) &  5.13(-3) &  2.94(-3) &  8.3(-4) &  2.2(-4) &  3.0(-5) \\
  2 & 0.135     &  1.16(-2) &  8.69(-3) &  5.99(-3) &  2.48(-3) &  9.5(-4) &  2.3(-4) \\
  0 & 1.0       &  2.71(-2) &  2.31(-2) &  1.85(-2) &  1.07(-2) &  5.81(-3) &  2.31(-3) \\
 -2 & 7.4       &  1.27(-1) &  1.18(-1) &  1.06(-1) &  7.91(-2) &  5.67(-2) &  3.37(-2) \\
 -4 & 54.6      &  6.12(-1) &  5.98(-1) &  5.78(-1) &  5.26(-1) &  4.69(-1) &  3.88(-1) \\
 -6 & 403.4     &  9.99(-1) &  9.98(-1) &  9.97(-1) &  9.93(-1) &  9.88(-1) &  9.77(-1) \\
\hline\\[-4pt]
\multicolumn{8}{c}{$\Delta J^2_0/\bar{J}^0_0$} \\[4pt]
\hline\\[-4pt]
 & & \multicolumn{6}{c}{$k$} \\
z & \multicolumn{1}{c}{$\tau$} & 0.1 & 0.2& 0.3 & 0.5 & 0.7 & 1.0 \\
\hline\\[-4pt]
  8 & 0.00033   &  1.61(-1) &  1.27(-1) &  7.26(-2) &  1.54(-2) &  2.52(-3) &  1.4(-4) \\
  6 & 0.0025    &  1.50(-1) &  1.39(-1) &  9.65(-2) &  3.06(-2) &  7.49(-3) &  7.9(-4) \\
  4 & 0.018     &  1.24(-1) &  1.33(-1) &  1.11(-1) &  5.25(-2) &  1.93(-2) &  3.65(-3) \\
  2 & 0.135     &  7.14(-2) &  8.66(-2) &  8.70(-2) &  6.14(-2) &  3.38(-2) &  1.16(-2) \\
  0 & 1.0       &  9.32(-3) &  1.53(-2) &  2.02(-2) &  2.33(-2) &  1.99(-2) &  1.24(-2) \\
 -2 & 7.4       & -5.4(-4) & -3.2(-4) & -1.1(-5) &  6.2(-4) &  1.08(-3) &  1.35(-3) \\
\hline\\[-4pt]                                             
\multicolumn{8}{c}{$\Delta \tilde{J}^2_1/\bar{J}^0_0$} \\[4pt]
\hline\\[-4pt]
 & & \multicolumn{6}{c}{$k$} \\
z & \multicolumn{1}{c}{$\tau$} & 0.1 & 0.2& 0.3 & 0.5 & 0.7 & 1.0 \\
\hline\\[-4pt]
  8 & 0.00033  & -1.09(-1) & -9.63(-2) & -5.68(-2) & -1.23(-2) & -2.02(-3) & -1.2(-4) \\
  6 & 0.0025   & -9.55(-2) & -1.03(-1) & -7.46(-2) & -2.43(-2) & -6.02(-3) & -6.4(-4) \\
  4 & 0.018    & -7.20(-2) & -9.50(-2) & -8.43(-2) & -4.14(-2) & -1.54(-2) & -2.97(-3) \\
  2 & 0.135    & -3.85(-2) & -6.11(-2) & -6.62(-2) & -4.89(-2) & -2.73(-2) & -9.47(-3) \\
  0 & 1.0      & -8.38(-3) & -1.54(-2) & -1.99(-2) & -2.16(-2) & -1.80(-2) & -1.10(-2) \\
 -2 & 7.4      & -2.6(-4) & -5.3(-4) & -7.9(-4) & -1.23(-3) & -1.49(-3) & -1.57(-3) \\
\hline\\[-4pt]
\multicolumn{8}{c}{$\Delta \tilde{J}^2_2/\bar{J}^0_0$} \\[4pt]
\hline\\[-4pt]
 & & \multicolumn{6}{c}{$k$} \\ 
z & \multicolumn{1}{c}{$\tau$} & 0.1 & 0.2& 0.3 & 0.5 & 0.7 & 1.0 \\
\hline\\[-4pt]
  8 & 0.00033  & -4.39(-2) & -4.23(-2) & -2.58(-2) & -5.75(-3) & -9.7(-4) & -5.4(-5) \\
  6 & 0.0025   & -3.60(-2) & -4.35(-2) & -3.30(-2) & -1.12(-2) & -2.85(-3) & -3.1(-4) \\
  4 & 0.018    & -2.34(-2) & -3.70(-2) & -3.54(-2) & -1.86(-2) & -7.19(-3) & -1.46(-3) \\
  2 & 0.135    & -9.27(-3) & -2.03(-2) & -2.51(-2) & -2.09(-2) & -1.24(-2) & -4.59(-3) \\
  0 & 1.0      & -1.59(-3) & -4.56(-3) & -7.15(-3) & -9.20(-3) & -8.22(-3) & -5.39(-3) \\
 -2 & 7.4      & -3.8(-5) & -1.4(-4) & -2.7(-4) & -5.5(-4) & -7.5(-4) & -8.6(-4) \\
\hline\\[-4pt]
\end{tabular}
\end{table}


%

\begin{figure}
\centerline{\epsfig{figure=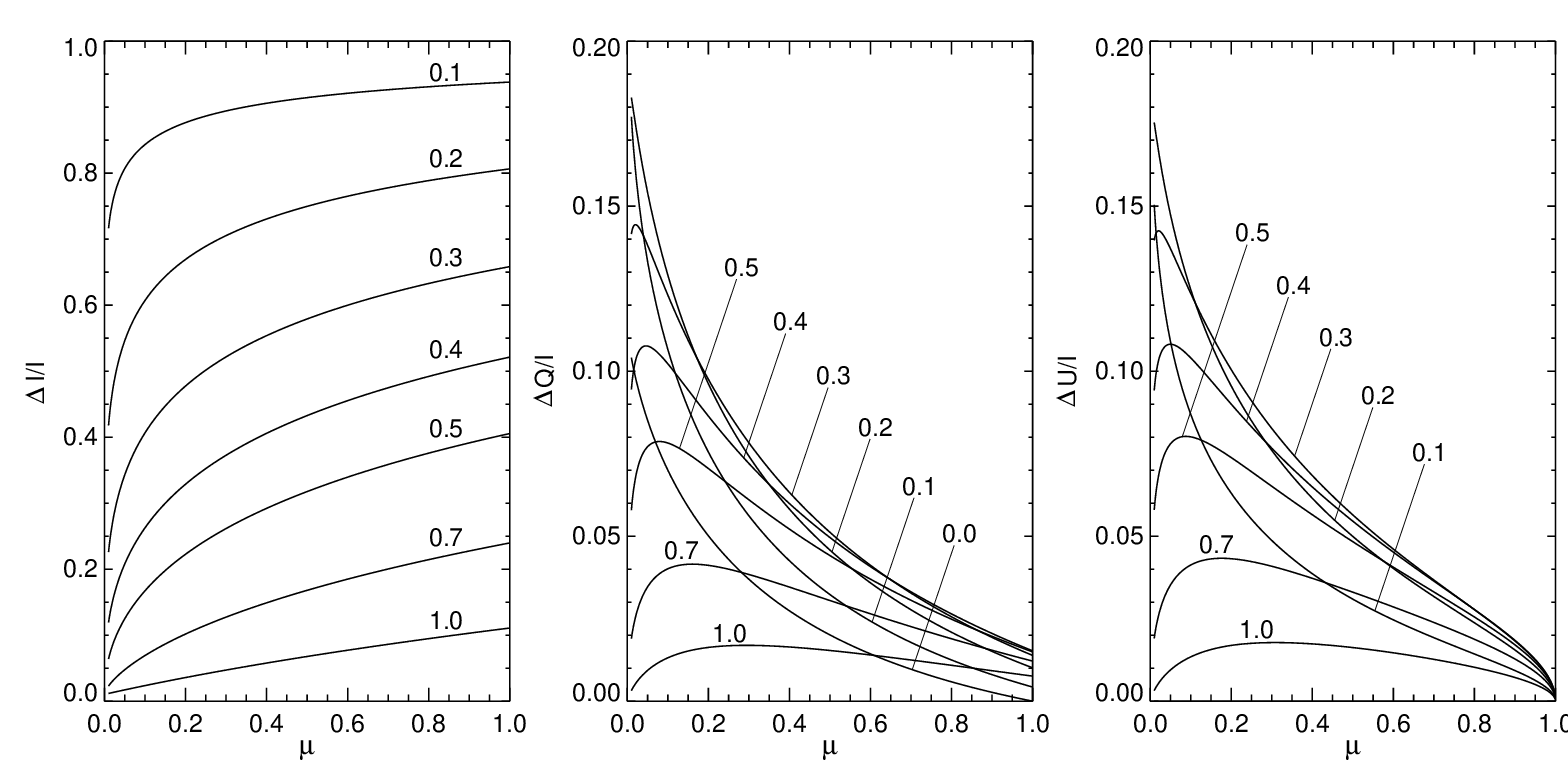, width=16cm}}
\caption{Center-to-limb variations of amplitudes $\Delta_1I/\bar{I}$,
$\Delta_1Q/\bar{I}$, and $\Delta_2U/\bar{I}$ corresponding to the amplitudes
in Fig.~(\ref{fig07}), for observation centered and {\em along the slabs}, i.e., 
at $x=0$, and along the $y$ axis (LOS with $\chi=0$). Labels indicate wavenumber $k$.\label{fig08}}
\end{figure}

For symmetry reasons (see Sect.~\ref{sect24}), 
$\delta\hat{J}^2_1=\delta\hat{J}^2_2=0$ and therefore, $\delta\hat{S}^2_1=\delta\hat{S}^2_2=0$;
moreover, $\Delta_1\tilde{J}^2_1=\Delta_2\tilde{J}^2_2=0$ (see Sect.~\ref{sect43})
and therefore, $\Delta_1\tilde{S}^2_1=\Delta_2\tilde{S}^2_2=0$.

The behavior of $\Delta J^0_0$ and $\Delta S^0_0$ is only slightly affected by polarization and it
closely resembles the behavior of the mean intensity and source function fluctuations of 
the unpolarized radiation problem (cf. Kneer 1981).

On the other hand, the amplitudes $\Delta J^2_0$, $\Delta\tilde{J}^2_1$, and $\Delta\tilde{J}^2_2$
(hence, the corresponding $\Delta S^2_0$, $\Delta\tilde{S}^2_1$, and $\Delta\tilde{S}^2_2$ ones), 
share the following asymptotic behaviors with depth ($z\rightarrow -\infty$), and close to the surface ($z\rightarrow \infty$).
Deeper than the thermalization depth, the radiation field becomes isotropic and therefore, 
$\Delta J^2_0, \Delta\tilde{J}^2_1, \Delta\tilde{J}^2_2 \rightarrow 0$;
towards the surface, horizontal transfer smears out 
horizontal fluctuations, thus recovering the unperturbed limit and therefore,
$\Delta J^2_0, \Delta\tilde{J}^2_1, \Delta\tilde{J}^2_2 \rightarrow 0$
(but for $k=0$, which corresponds to the planeparallel limit).
Between these two limits, the amplitudes rise in absolute value, reach a maximum at some height,
and decrease again.
This maximum absolute value tends to zero as $k\rightarrow\infty$.
The reason is that for increasingly small structures, they {\em uniformly} become optically thin,
approaching the planeparallel unperturbed limit 
($\Delta J^2_Q\rightarrow 0$) {\em throughout the whole atmosphere}.

It is interesting to study the behavior with wavenumber $k$ at a given height, for example,
at $z\approx 2.3$, which corresponds to $\tau_{\nu_0}\approx 1$ at $\mu=0.1$ (i.e., close to the limb). 
The amplitudes monotonically increase (in absolute value) with wavenumber up to $k\sim 0.2$;
beyond that, $\Delta J^2_Q$ and $\Delta S^2_Q$ become smaller
the larger the horizontal wavenumber $k$
---which will explain the behavior of the emergent fractional polarization (see below).
When horizontal inhomogeneities appear, they break the rotational symmetry of the radiation field, 
which perturb the (vertical) anisotropy $\Delta J^2_0$ and  yields $\tilde{J}^2_1$, $\tilde{J}^2_2\ne 0$. 
Then, at some point, those horizontal fluctuations become optically thin at the `height of formation'
and finally, the radiation field recovers its original symmetry.

Figure~\ref{fig08} shows the center-to-limb variation {of the  
Stokes profiles amplitudes} when observing along the `slabs', i.e., along the $y$-axis.
Due to symmetry reasons, for observations along the invariant direction, $I$ and $Q$ fluctuate cosinusoidally, (i.e., $\Delta_2I=\Delta_2Q=0$) just like the perturbation, while $U$ fluctuates sinusoidally (i.e., $\Delta_1U=0$).
This can also be understood by taking $\chi=90^\circ$ ($\lambda=0$) in Eqs.~(\ref{eq25})-(\ref{eq26}).
Then, amplitudes $\Delta_1$ and $\Delta_2$ are independent between them, and the just mentioned result
follows from Eqs.~(\ref{eq02}) evaluated at $\chi=90^\circ$, and the fact that
the only non vanishing components are $\Delta S^0_0$, $\Delta S^2_0$, $\Delta \tilde{S}^2_1$, 
and $\Delta \tilde{S}^2_2$.

The limb darkening of the $\Delta_1I$ amplitudes adds up to the limb darkening of the mean model.
At large $k$, however, the fluctuations become optically thin, $\Delta I\rightarrow 0$ at all $\mu$,
and the limb darkening law of the unperturbed model is recovered.
Amplitudes $\Delta_1Q$ and $\Delta_2U$ increase towards the limb before steeply falling to zero 
at $\mu\rightarrow 0$; no horizontal transfer effects exist in the limit of tangential observation. 
This is a consequence of the amplitudes $\Delta S^2_Q$ falling towards zero
when approaching the surface.  
The non monotonic behavior of $\Delta S^2_Q$ with varying wavenumber $k$ at a fixed height
manifests here as an increase and decrease of $Q$ and $U$ with $k$ at a given $\mu$.

\subsection{Scattering line polarization with source function fluctuations}\label{sect61}

The main difference between coherent and resonance scattering is that in the latter case, photons may redistribute (here, completely) within the spectral profile. As a consequence, the thermalization depth is shallower in the former case (e.g., Mihalas 1978), which leads to steeper gradients of the source function and hence, to larger radiation field anisotropies and emergent fractional polarization. 
As it turns out, horizontal transfer effects ---and hence, $\Delta J^2_Q$--- are also reduced in the resonance case.

\begin{figure}
\centerline{\epsfig{figure=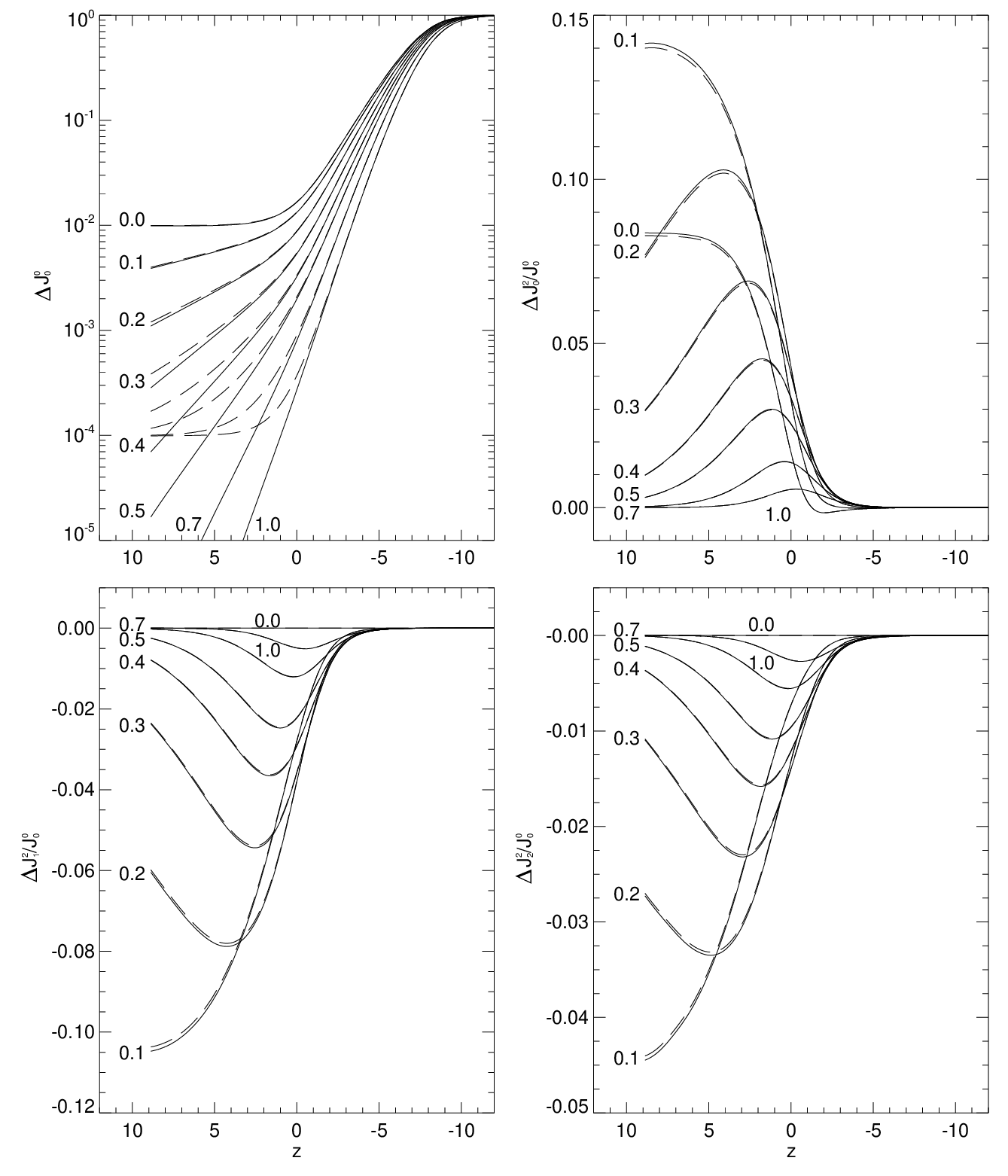, width=16cm}}
\caption{Scattering polarization in an atmosphere with horizontal, cosinusoidal 
fluctuations of the Planck function (Eq.~(\ref{eq21})). Labels indicate wavenumber $k$. 
A $0\rightarrow 1$ transition with $\epsilon=10^{-4}$ and no depolarizing collisions
is considered.
From top to bottom, and from left to right: $\Delta J^0_0$, $\Delta J^2_0/\bar{J}^0_0$,
$\Delta \tilde{J}^2_1/\bar{J}^0_0$, $\Delta \tilde{J}^2_2/\bar{J}^0_0$, respectively 
(solid lines), and $\Delta S^0_0$, $\Delta S^2_0/\bar{S}^0_0$,
$\Delta \tilde{S}^2_1/\bar{S}^0_0$, $\Delta \tilde{S}^2_2/\bar{S}^0_0$ (dashed lines).\label{fig04}}
\end{figure}

We consider resonance scattering in a constant properties, semi-infinite atmosphere with sinusoidal, horizontal fluctuations of the Planck function. In particular, we consider an infinitely strong line ($\kappa^{\rm cont}=0$). 
The line opacity exponentially stratified with height. We normalize spatial scales to the opacity scale height ($\kappa^{\rm line}=\kappa^{\rm line}_\circ{\rm e}^{-z}$), and an arbitrary height reference is set by choosing $\kappa^{\rm line}_\circ=1$.
We consider a Gaussian absorption profile (thermal motions and microturbulence dominate), and
we normalize frequencies to the thermal Doppler width ($v=(\nu-\nu_0)/\Delta\nu_D$; $\phi_v={\rm e}^{-v^2}/\sqrt{\pi}$).
The optical depth {at the line center}, taken from the surface ($z\rightarrow\infty$), at a given $\mu$ is then $\tau={\rm e}^{-z}/\sqrt{\pi}\mu$; conversely, the height (in units of the opacity scale) at which $\tau=1$ along a LOS with $\mu$ is $z=\log(1/\mu\sqrt{\pi})$. 
Finally, we normalize $\Delta B$, $\Delta J^0_0$, $\Delta S^0_0$ amplitudes to the mean Planck function.

Figure~\ref{fig04} and Table~\ref{tab3} show 
the solution of this transfer problem for a strong scattering line ($\kappa^{\rm cont}=0$, $\epsilon=10^{-4}$),
and a Planck function fluctuation $\Delta B/\bar{B}=1$.
They show the variation of $\Delta J^0_0$, $\Delta J^2_0$, $\Delta\tilde{J}^2_1$, and $\Delta\tilde{J}^2_2$, 
and the corresponding $\Delta S^0_0$, $\Delta S^2_0$, $\Delta\tilde{S}^2_1$, and $\Delta\tilde{S}^2_2$,
with height as a function of the horizontal wavenumber $k$.
Clearly, the linearity and symmetry considerations given in Sect.~(\ref{sect62}) apply here too.

As in the coherent case, $\Delta J^0_0$ and $\Delta S^0_0$ are only slightly affected by polarization if compared with the solution of the unpolarized problem ($\delta\rightarrow\infty$).
As seen by the $k=0$ limit in Figs.~\ref{fig07} and \ref{fig04}, the 2-level atom thermalizes deeper in the atmosphere (Mihalas 1978). 
The shallower gradient of the source function leads to a smaller anisotropy ($J^2_0$; e.g., Manso Sainz 2002), and to smaller horizontal anisotropies ($\Delta J^2_Q$).
Otherwise, the behavior of $\Delta J^2_Q$ with height is identical in the coherent and CRD case.

\begin{table}
\caption{Radiation field tensor amplitudes for $\epsilon=10^{-4}$\label{tab3}}
\begin{tabular}{r.......}
\hline\\[-4pt]
\multicolumn{8}{c}{$\Delta J^0_0/B$} \\[4pt]
\hline\\[-4pt]
 & & \multicolumn{6}{c}{$k$} \\
z & \multicolumn{1}{c}{$\tau$} & 0.1 & 0.2& 0.3 & 0.5 & 0.7 & 1.0 \\
\hline\\[-4pt]
 8 & 0.00033 & 4.21(-3) & 1.30(-3) & 3.64(-4) & 2.62(-5) & 1.5(-6) &           \\
 6 & 0.0025  & 5.07(-3) & 1.91(-3) & 6.55(-4) & 7.24(-5) & 8.4(-6) &  1.9(-7) \\
 4 & 0.018   & 6.18(-3) & 2.84(-3) & 1.19(-3) & 1.99(-4) & 3.65(-5) &  4.7(-6) \\
 2 & 0.135   & 8.01(-3) & 4.45(-3) & 2.27(-3) & 5.68(-4) & 1.57(-4) &  3.51(-5) \\
 0 & 1.0     & 1.33(-2) & 8.72(-3) & 5.34(-3) & 1.94(-3) & 7.70(-4) &  2.63(-4) \\
-2 & 7.4     & 3.50(-2) & 2.62(-2) & 1.86(-2) & 9.22(-3) & 4.87(-3) &  2.27(-3) \\
-4 & 54.6    & 1.13(-1) & 9.43(-2) & 7.61(-2) & 4.87(-2) & 3.22(-2) &  1.92(-2) \\
-6 & 403.4   & 3.33(-1) & 3.04(-1) & 2.71(-1) & 2.14(-1) & 1.70(-1) &  1.25(-1) \\
\hline\\[-4pt]
\multicolumn{8}{c}{$\Delta J^2_0/\bar{J}^0_0$} \\[4pt]
\hline\\[-4pt]
 & & \multicolumn{6}{c}{$k$} \\
z & \multicolumn{1}{c}{$\tau$} & 0.1 & 0.2& 0.3 & 0.5 & 0.7 & 1.0 \\
\hline\\[-4pt]
 8 & 0.00033 & 1.41(-1) & 8.33(-2) & 3.50(-2) & 4.34(-3) & 4.95(-4) & 3.9(-5) \\
 6 & 0.0025  & 1.36(-1) & 9.66(-2) & 4.94(-2) & 9.17(-3) & 1.55(-3) & 1.4(-4) \\
 4 & 0.018   & 1.22(-1) & 1.03(-1) & 6.38(-2) & 1.77(-2) & 4.43(-3) & 6.6(-4) \\
 2 & 0.135   & 8.89(-2) & 8.90(-2) & 6.74(-2) & 2.78(-2) & 1.02(-2) & 2.56(-3) \\
 0 & 1.0     & 3.29(-2) & 4.29(-2) & 4.08(-2) & 2.53(-2) & 1.34(-2) & 5.43(-3) \\
-2 & 7.4     & 1.92(-3) & 5.87(-3) & 7.87(-3) & 7.67(-3) & 5.76(-3) & 3.46(-3) \\
\hline\\[-4pt]                                             
\multicolumn{8}{c}{$\Delta \tilde{J}^2_1/\bar{J}^0_0$} \\[4pt]
\hline\\[-4pt]
 & & \multicolumn{6}{c}{$k$} \\
z & \multicolumn{1}{c}{$\tau$} & 0.1 & 0.2& 0.3 & 0.5 & 0.7 & 1.0 \\
\hline\\[-4pt]
 8 & 0.00033 & -1.04(-1) & -6.52(-2) & -2.78(-2) & -3.47(-3) & -3.8(-4) & -1.1(-5) \\
 6 & 0.0025  & -9.72(-2) & -7.48(-2) & -3.91(-2) & -7.36(-3) & -1.24(-3) & -1.1(-4) \\
 4 & 0.018   & -8.33(-2) & -7.85(-2) & -5.03(-2) & -1.42(-2) & -3.60(-3) & -5.5(-4) \\
 2 & 0.135   & -5.95(-2) & -6.81(-2) & -5.33(-2) & -2.25(-2) & -8.38(-3) & -2.15(-3) \\
 0 & 1.0     & -2.80(-2) & -3.79(-2) & -3.56(-2) & -2.18(-2) & -1.16(-2) & -4.75(-3) \\
-2 & 7.4     & -5.54(-3) & -8.58(-3) & -9.43(-3) & -8.00(-3) & -5.76(-3) & -3.41(-3) \\
\hline\\[-4pt]
\multicolumn{8}{c}{$\Delta \tilde{J}^2_2/\bar{J}^0_0$} \\[4pt]
\hline\\[-4pt]
 & & \multicolumn{6}{c}{$k$} \\
z & \multicolumn{1}{c}{$\tau$} & 0.1 & 0.2& 0.3 & 0.5 & 0.7 & 1.0 \\
\hline\\[-4pt]
 8 & 0.00033 & -4.33(-2) & -2.92(-2) & -1.28(-2) & -1.61(-3) & -1.5(-4) &           \\
 6 & 0.0025  & -3.89(-2) & -3.27(-2) & -1.77(-2) & -3.45(-3) & -5.8(-4) & -2.7(-5) \\
 4 & 0.018   & -3.07(-2) & -3.29(-2) & -2.22(-2) & -6.60(-3) & -1.73(-3) & -2.7(-4) \\
 2 & 0.135   & -1.87(-2) & -2.62(-2) & -2.22(-2) & -1.02(-2) & -4.01(-3) & -1.12(-3) \\
 0 & 1.0     & -7.30(-3) & -1.32(-2) & -1.41(-2) & -9.66(-3) & -5.53(-3) & -2.53(-3) \\
-2 & 7.4     & -1.44(-3) & -3.09(-3) & -3.91(-3) & -3.80(-3) & -2.97(-3) & -1.98(-3) \\
\hline\\[-4pt]

\end{tabular}
\end{table}

It is interesting to study the behavior with wavenumber $k$ at a given height, say,
$z\approx 1.73$, which corresponds to $\tau_{\nu_0}\approx 1$ at $\mu=0.1$ (i.e., close to the limb). 
The amplitudes increase (in absolute value) monotonically with wavenumber up to $k\sim 0.2$;
beyond that, $\Delta J^2_Q$ and $\Delta S^2_Q$ become smaller
the larger the horizontal wavenumber $k$
---which will explain the behavior of the emergent profiles seen in Fig.~(\ref{fig05}) (see below).
First, horizontal inhomogeneities perturb the (vertical) anisotropy $\Delta J^2_0$,
breaking the rotational symmetry of the radiation field, which yields
$\tilde{J}^2_1$, $\tilde{J}^2_2\ne 0$. Then, at some point, horizontal
fluctuations become optically thin at the `height of formation'
and finally, the radiation field recovers its original symmetry.

\begin{figure}
\centerline{\epsfig{figure=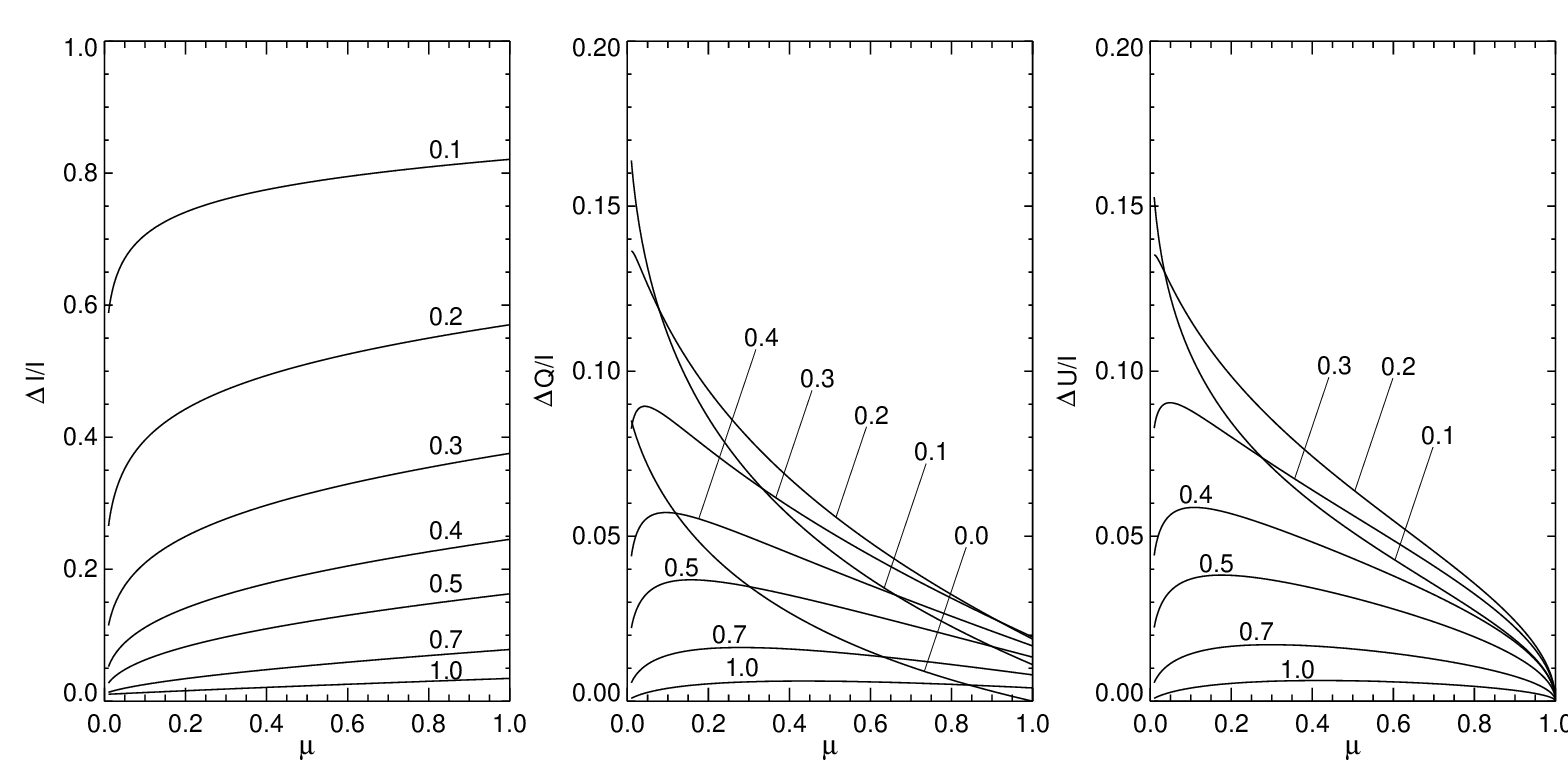, width=16cm}}
\caption{Center-to-limb variations of amplitudes $\Delta_1I/\bar{I}$,
$\Delta_1Q/\bar{I}$, and $\Delta_2U/\bar{I}$ corresponding to the amplitudes
in Fig.~(\ref{fig04}), for observation {\em centered and along the slabs}, i.e., 
at $x=0$, and along the $y$ axis (LOS with $\chi=0$).
Labels indicate wavenumber $k$.\label{fig06}}
\end{figure}

From the statistical tensors amplitudes $\Delta S^K_Q$ we compute the amplitudes of the emergent polarization profiles for any direction.
In particular, Fig.~\ref{fig06} shows the center-to-limb variation of the line core 
Stokes profiles amplitudes when observing along the `slabs', i.e., along the $y$-axis.
As in Sect.~\ref{sect62}, the behavior with $k$ of the emergent Stokes parameters amplitudes at a given $\mu$ can be understood from the 
behavior with wavenumber of  $\Delta S^K_Q$ at a given height (remembering that, for example, $z\approx 1.73$  corresponds to $\tau_{\nu_0}\approx 1$ at $\mu=0.1$).

Due to symmetry reasons, for observations along the invariant direction, $I$ and $Q$ fluctuate cosinusoidally, (i.e., $\Delta_2I=\Delta_2Q=0$) just like the perturbation, while $U$ fluctuates sinusoidally (i.e., $\Delta_1U=0$).
This can also be understood by taking $\chi=90^\circ$ ($\lambda=0$) in Eqs.~(\ref{eq25})-(\ref{eq26}).
Then, amplitudes $\Delta_1$ and $\Delta_2$ are independent between them, and the just mentioned result
follows from Eqs.~(\ref{eq02}) evaluated at $\chi=90^\circ$, and the fact that
the only non vanishing components are $\Delta S^0_0$, $\Delta S^2_0$, $\Delta \tilde{S}^2_1$, 
and $\Delta \tilde{S}^2_2$.

\begin{figure}
\centerline{\epsfig{figure=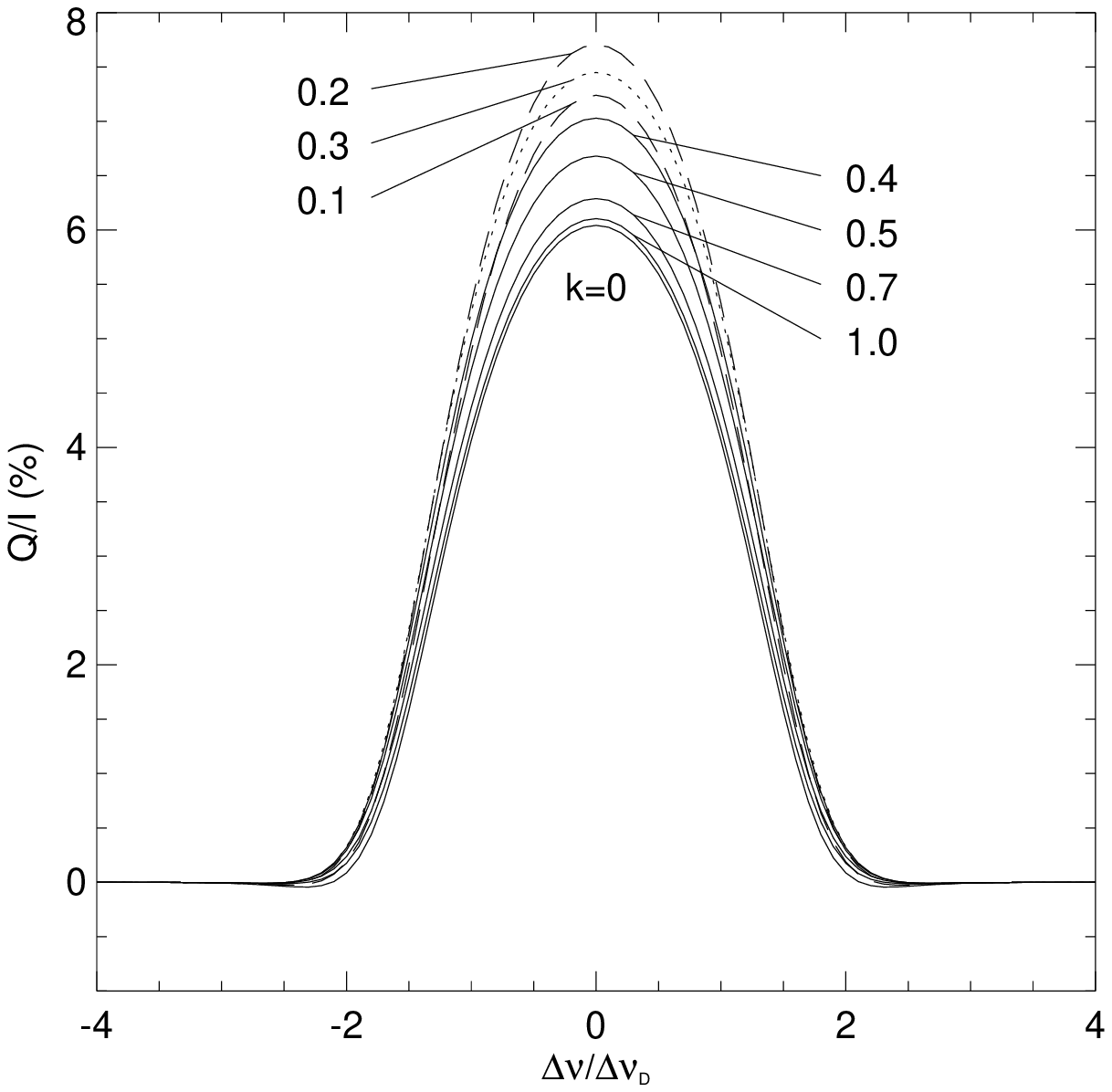, width=8cm}\epsfig{figure=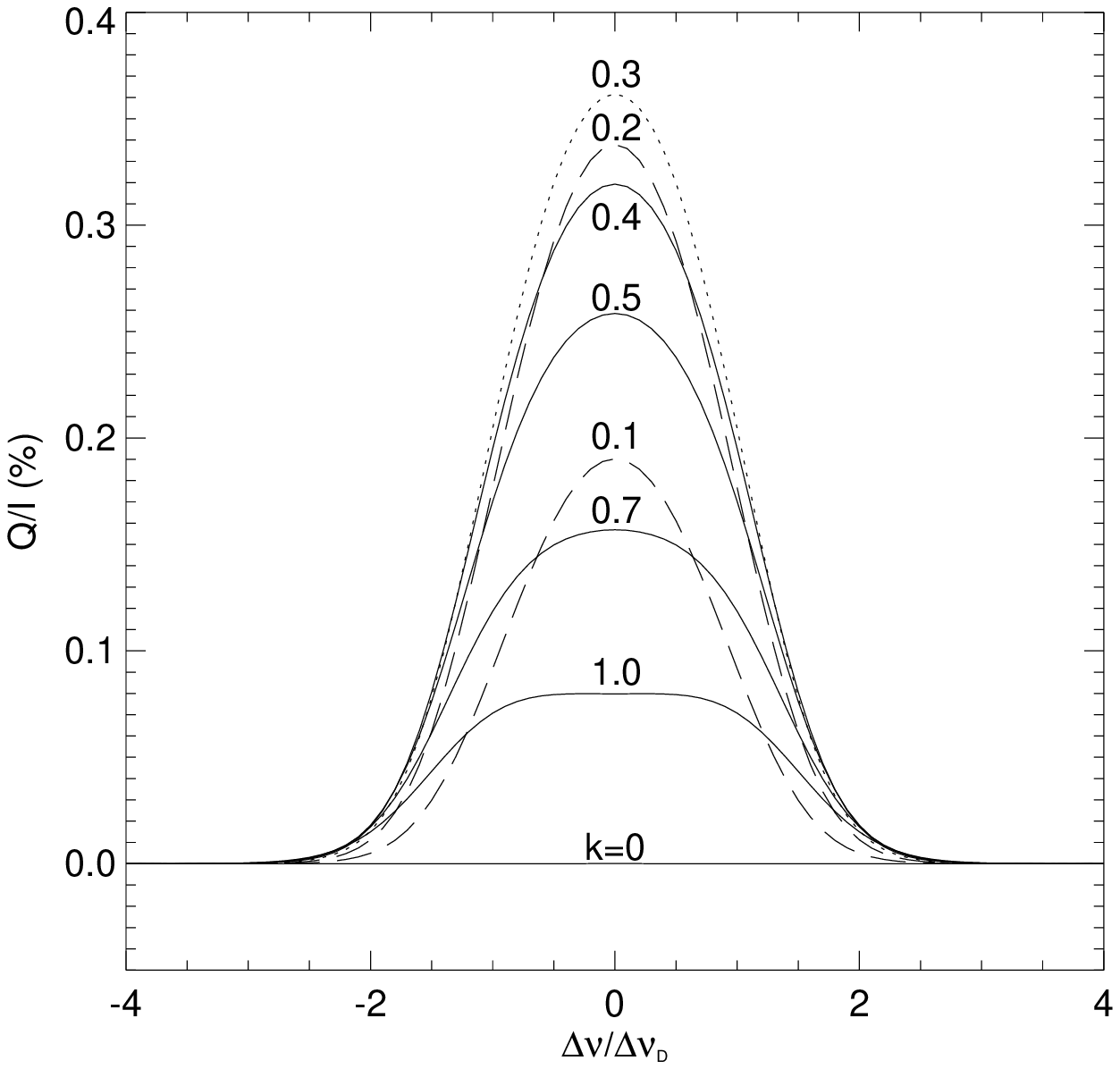, width=8cm}}
\caption{\label{fig05}Emergent fractional polarization $Q/I$ close to the limb at $\mu=0.1$ (left), 
and at disk center (right) from an atmosphere with a fluctuating Planck function 
$B=\bar{B}[1+0.2\cos(kx)]$; labels indicate wavenumber $k$.
The LOS in the left panel is taken at $x=0$ and along the slabs
(i.e., along the $y$-axis, see Fig.~\ref{fig01}), and the positive direction of $Q$ is perpendicular to 
the limb. The line of sight in the right panel is taken at $x=0$ and the positive-$Q$ 
direction is perpendicular to the slab.}
\end{figure}

Figure~\ref{fig05} shows the full profile of the emergent fractional polarization for a specific fluctuation with $\Delta B/\bar{B}=0.2$.
This has been calculated using the linearity of the problem, from the amplitudes of the Stokes profiles  of the 
planeparallel case ($k=0$), and 0.2 times the amplitude of the sinusoidally perturbed model.  
When observing at $\mu=0.1$ and along the $y$-axis (left panel), the amount of polarization increases
with wavenumber for $k<0.2$; at $k\sim 0.2$ it changes trend, and for larger $k$ values the plane-parallel
limit is again recovered. 
A similar behavior is observed at disk center (right panel), where the presence of horizontal fluctuations alone
{breakes} the symmetry of the problem and is able to generate a polarization signal. 
Since there is more radiation along the slabs than perpendicularly to them, the polarization
direction is perpendicular to the slab (i.e., along the $x$-axis).

\subsection{Scattering line polarization, source function and opacity fluctuations}\label{sect63}

We consider now the additional effect of an opacity fluctuation.
If the opacity and Planck function are in phase, hotter regions are more opaque, cooler regions more transparent, and
radiation is {\em channeled} from brighter to fainter regions (Cannon 1970). 
If, on the contrary, opacity and Planck function fluctuate in anti-phase, hot regions are more transparent (hence, brighter), emission in cooler regions is further blocked (thus fainting), and contrast is enhanced. 

The effect of these mechanisms on the emergent polarization is illustrated in Fig.~\ref{fig09}.
It shows the effect of opacity fluctuations in phase ($\alpha=+0.1$), or in anti-phase ($\alpha=-0.1$) 
with the Planck function, for fluctuations with $k=0.3$, and $\Delta B/\bar{B}=0.2$ (as in Fig.~\ref{fig05}.
The modification of the $Q/I$ profiles by such opacity inhomogeneities is negligible ($\sim$0.1\%), 
compared with  the intrinsic polarization from the unperturbed model and the Planck function fluctuation itself
in close-to-the-limb observations ($\mu=0.1$).
More interesting is the polarization signal at disk center ($\mu=1$), 
which is exclusively due to the horizontal inhomogeneities.
Opacity fluctuations with $\alpha>0$ ($\alpha<0$) enhance (diminish) the polarization signal generated by the Planck function fluctuations. 

Close to the limb, 
anti-phase fluctuations ($\alpha<0$) hamper blurring of horizontal inhomogeneities, which enhances Stokes $U$ at the edge between hot and cool slabs ($x=P/4$).
Fluctuations in phase enhance radiation channeling, hence horizontal blurring, and Stokes $U$ decreases. 
At disk center ($\mu=1$), $U=0$ for symmetry reasons (see Fig.~2).

\begin{figure}
\centerline{\epsfig{figure=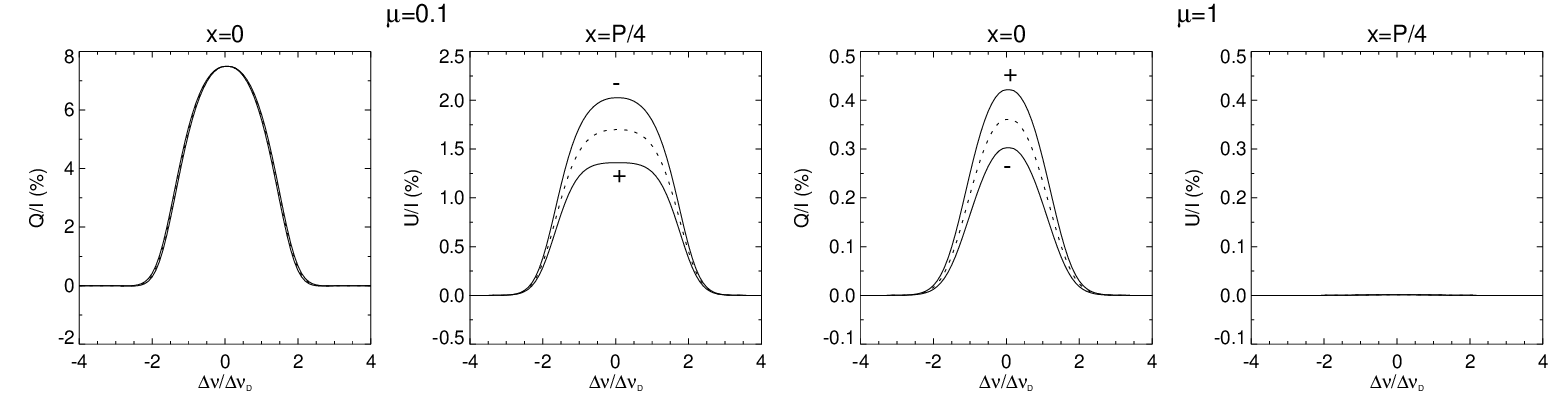, width=\textwidth}}
\caption{
Emergent fractional polarization from an atmosphere with horizontal fluctuations of the 
Planck function and opacity.
The wavenumber of the fluctuations is $k=0.3$, the Planck function amplitude is $\Delta B/\bar{B}=0.2$,
and the opacity fluctuations are $\alpha=+0.1$ (solid lines labeled with a $+$ sign),
$-0.1$ (solid lines labeled with a $-$ sign), or 0 (dotted lines).
Leftmost panels correspond to a close-to-the-limb ($\mu=0.1$) LOS; 
rightmost panels to a disc center ($\mu=1$) geometry.
$Q/I$ is shown at $x=0$ with a LOS along the $y$-axis; 
$U/I$ is shown at $x=P/4$ ($P=2\pi/k$ is the period of the horizontal perturbation).
\label{fig09}
}
\end{figure}

\subsection{Hanle effect with source function and opacity fluctuations}\label{sect64}

In the presence of a magnetic field, the problem loses all its symmetries and all twelve unknown amplitudes $\Delta_{1, 2}S^K_Q$ must be considered.
Yet, there are three important cases in which the field is symmetric enough to preserve some symmetries, which lead to simplifications of the description and are interesting for understanding the radiative transfer problem.
We will study in some detail the following three configurations: vertical magnetic field, and horizontal magnetic field aligned with, or transversal to, the thermodynamic fluctuations.
The simplifications to which these symmetric configurations lead are summarized in Table~\ref{tab04}.

It is important to note that $\Delta_{1, 2}S^K_Q\rightarrow 0$ (for $k\ne 0$) when approaching to the stellar surface, regardless of the magnetic field configuration. 
Consequently, $\Delta_{1, 2}S_I, \Delta_{1, 2}S_Q, \Delta_{1, 2}S_U \rightarrow 0$, and there are no observable multidimensional effects in the limit of tangential observation. 
In the following, we shall consider the case of finite $\mu>0$ observations.

As is well known, a vertical field produces no Hanle effect in a plane-parallel medium (just consider $\sin\theta_B=0$ and $J^2_{1}=J^2_2=0$ in Eqs.~(\ref{eq10})).
This is no longer true in the presence of horizontal inhomogeneities, since the vertical direction is no longer a symmetry axis for the radiation field.
Symmetry considerations and Eqs.~(\ref{eq10}) for the amplitudes $\Delta_{1, 2}S^2_Q$ show that the only non vanishing amplitudes in this case are the ones listed in Table~(\ref{tab04}).
Moreover, when the magnetic field is strong enough to reach saturation ($\Gamma\rightarrow\infty$), the only non vanishing amplitude is  $\Delta_1S^2_0$.
Figure~\ref{fig10} shows $\Delta Q/\bar{I}$ and $\Delta U/\bar{I}$ emergent at $x=0$ (where the fluctuation of the Planck function is maximum), for a ray along the $y$-axis and $\mu=0.1$ (close-to-the-limb observation), or with $\mu=1$ (disk center observation or forward scattering geometry). 
Upper panels show that the effect of a vertical magnetic field is a depolarization of the signal produced by the horizontal fluctuations, and a rotation of the polarization plane. 
For strong fields ($\Gamma\gg 1$), depolarization is partial in the $\mu=0.1$ geometry, and complete at $\mu=1$.

When the magnetic field is horizontal and aligned with the LOS (central panels), the effect when observing at $\mu=0.1$ can be interpreted as in the classical picture in a planeparallel atmosphere (a depolarization, which may be total, and rotation of the polarization plane).
In forward scattering geometry, the field destroys the orthogonal polarization generated by the horizontal inhomogeneities, while no rotation of the polarization plane is possible for symmetry reasons.

When the magnetic field is perpendicular to the LOS, no rotation of the polarization plane is possible at any $\mu$, due to symmetry. 
The magnetic field lies along the polarization direction which leads to partial depolarization at $\mu=0.1$ due to the precession of the LOS component of the dipole.
In forward scattering ($\mu=1$), however, overpolarization is possible because the anisotropy is negative at $\tau\approx 1$.

All the previous discusions apply to spatially resolved observations.
It is however interesting to consider the effect of finite spatial resolution
by averaging the fluctuations over intervals $\Delta$ along the $x$ axis. 
Thus, the average amplitudes at $x=0$ are $\langle\delta I\rangle=\frac{1}{\Delta}
\int_{-\Delta/2}^{\Delta/2}\delta I {\rm d}x$, with analogous expressions for $\langle\delta Q\rangle$ and $\langle\delta U\rangle$.
The emergent fractional polarization with low spatial resolution at $x=0$ is then
\begin{equation}
\frac{\langle Q\rangle}{\langle I\rangle}=\frac{\bar{Q}(z_M)+\Delta_1Q(z_M) \;{\rm sinc}(\frac{k\Delta}{2})}{\bar{I}(z_M)+\Delta_1I(z_M)\; {\rm sinc}(\frac{k\Delta}{2})},
\qquad
\frac{\langle U\rangle}{\langle I\rangle}=\frac{\bar{U}(z_M)+\Delta_1U(z_M) \;{\rm sinc}(\frac{k\Delta}{2})}{\bar{I}(z_M)+\Delta_1I(z_M)\; {\rm sinc}(\frac{k\Delta}{2})},
\end{equation}
where $z_M$ is the outer free boundary of the atmosphere and ${\rm sinc}(x)=\sin(x)/x$.
At high spatial resolution, $\Delta\rightarrow 0$ and we recover previous results;
at low spatial resolution, $\Delta\rightarrow \infty$ and we recover the fractional polarization
$\bar{Q}/\bar{I}$ and $\bar{U}/\bar{I}$ of the average model. 
Actually, for $k\Delta/2\ge \pi$, the amplitudes fall below  20\% the resolved value, 
and for $k\Delta/2\ge 4\pi$ the amplitudes are always below 10\% the resolved value.
Therefore, for very low spatial resolutions the effects of horizontal fluctuations disappear and we recover the plane-parallel limit. 
In particular, if we have a large-scale magnetic field in an atmosphere with horizontal fluctuations of the thermodynamic parameters, in the limit of low resolution {only the Hanle effect of the large scale} magnetic field remains.

\begin{table}
\begin{center}
\caption{Independent unknowns according to the magnetic field geometry\label{tab04}}
\begin{tabular}{lc}
\hline
Geometry\footnotemark[1] & non-vanishing $\Delta_{1, 2}S^2_Q$ amplitudes\footnotemark[2] \\
\hline
No magnetic field ($B_x=B_y=B_z=0$)  & $\Delta_1S^2_0$, $\Delta_2\tilde{S}^2_1$, $\Delta_1\hat{S}^2_2$  \\
Vertical field ($B_x=B_y=0$)     & $\Delta_1S^2_0$, [$\Delta_2\tilde{S}^2_1$], [$\Delta_2\hat{S}^2_1$], [$\Delta_1\tilde{S}^2_2$], [$\Delta_1\hat{S}^2_2$] \\
Horizontal field across slabs ($B_y=B_z=0$)      & $\Delta_1S^2_0$, [$\Delta_2\tilde{S}^2_1$], [$\Delta_1\hat{S}^2_1$], $\Delta_1\tilde{S}^2_2$, [$\Delta_2\hat{S}^2_2$] \\    
Horizontal field along the slabs ($B_x=B_z=0$)       & $\Delta_{1}S^2_0$, [$\Delta_{2}S^2_0$], [$\Delta_{1, 2}\tilde{S}^2_1$], $\Delta_{1}\tilde{S}^2_2$, [$\Delta_{2}\tilde{S}^2_2$]\\    
\hline
\end{tabular}
\footnotetext[1]{Horizontal fluctuations $\sim\cos(kx)$ (invariant direction along $y$-axis).}
\footnotetext[2]{Elements between brackets vanish when the field saturates ($\Gamma\rightarrow\infty$).}
\end{center}
\end{table}

\begin{figure}
\centerline{\epsfig{figure=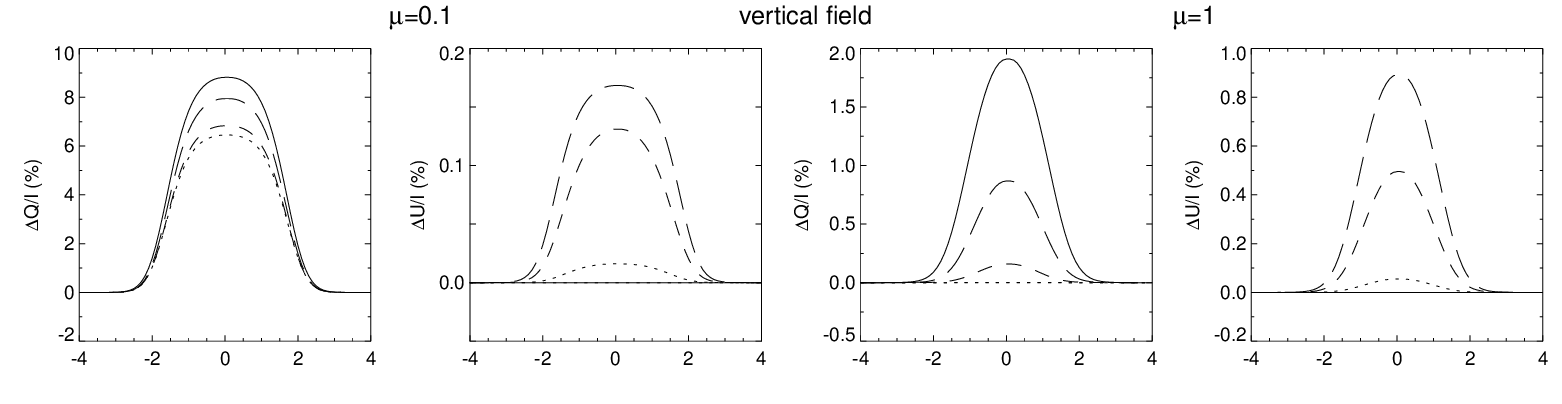, width=\textwidth}}
\centerline{\epsfig{figure=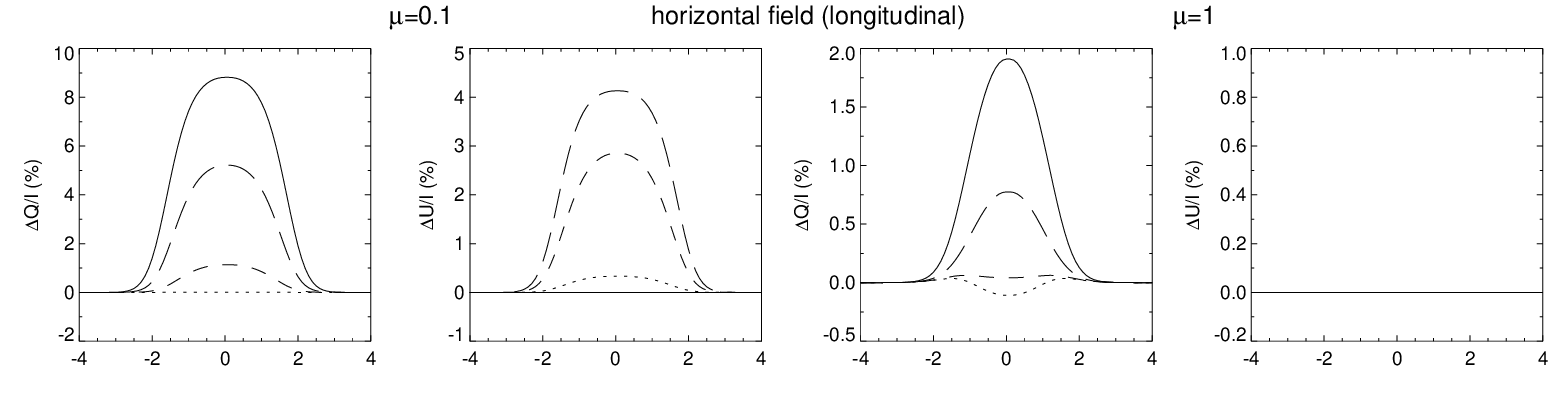, width=\textwidth}}
\centerline{\epsfig{figure=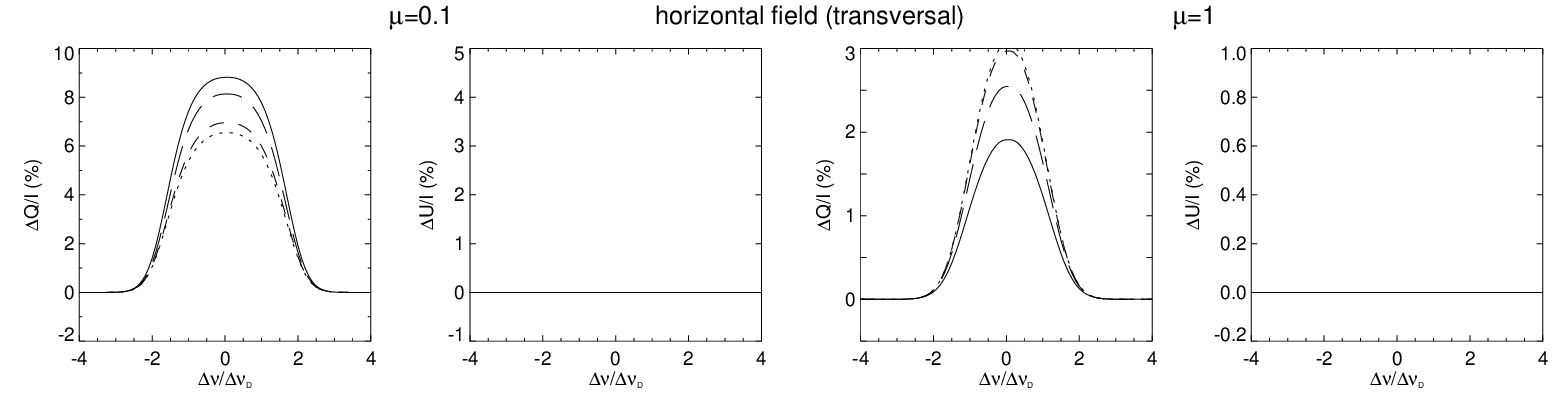, width=\textwidth}}
\caption{Amplitudes of the emergent Stokes parameters $\Delta Q/\bar{I}$ and $\Delta U/\bar{I}$ at 
$x=0$, observed along the $x$-axis at $\mu=0.1$ (left panels), and at $\mu=1$ (right panels).
A periodic horizontal fluctuation ($\sim\cos(0.3x)$) is assumed for the Planck function.
Top panels correspond to a vertical magnetic field; middle panels to a horizontal field
{along the $y$-axis; bottom panels to a horizontal field}
along the $x$-axis.
For a given magnetic field configuration the magnetic field strength is $\Gamma=0$ (solid line), $\Gamma=0.3$ (long-dashed), $\Gamma=1$ (dashed), and $\Gamma=10$ (dotted).
The $Q>0$ direction is parallel to the horizon on the left panels and 
transversal to the ``slab'' ($x$-axis) on the right ones.\label{fig10}
}
\end{figure}

\section{Conclusions}

In this paper we have formulated and solved the problem of scattering polarization in
horizontally inhomogeneous stellar atmospheres for continuum and resonance line radiation including the Hanle effect from random and deterministic magnetic fields.
If the amplitudes of the horizontal fluctuations of the Planck function and of the opacity are small enough (compared with the spatially-averaged values), the ensuing radiative transfer problem can be linearized;
furthermore, if the opacity does not fluctuate, the response of the radiation field to Planck function fluctuations is already linear. In both cases, two- and three-dimensional scattering polarization problems can be solved at the computational cost of one-dimensional problems, because no spatial discretization is needed along the horizontal directions, thus avoiding computationally costly interpolations between grid points. We have developed and implemented iterative numerical methods {and formal solvers of the transfer equations} to solve this type of multidimensional radiative transfer problems.

The numerical codes developed implementing these methods allow carrying out several interesting investigations. For instance, it is important to study the impact of the symmetry breaking effects caused by the presence of horizontal atmospheric inhomogeneities on the emergent scattering polarization and its importance relative to the modification of the linear polarization produced by the Hanle effect.  
Here we have given a first step towards such a goal by considering sinusoidal horizontal inhomogeneities in the Planck function and in the opacity, in the absence and in the presence of deterministic magnetic fields (i.e., with a  fixed orientation). The most important results of our study, to be taken into account when interpreting high-spatial resolution observations of the linear polarization produced by scattering processes in the solar atmosphere, are the following:


(1) When considering increasingly smaller horizontal atmospheric inhomogeneities (i.e., increasingly larger wavenumbers) we find that the amplitudes of the fractional scattering polarization signals first increase and then, beyond a given wavenumber, the inhomogeneities become optically thin and the planeparallel limit is recovered again (see Fig. 6). The maximum impact occurs for $k{\approx}0.2$ or $k{\approx}0.3$, which corresponds to P-values of about 20 opacity scale heights. Note that the symmetry breaking effects caused by the horizontal atmospheric inhomogeneities may produce forward-scattering $Q/I$ and $U/I$ signals, without the need of an inclined magnetic field.

(2) In the presence of atmospheric inhomogeneities with a dominant horizontal scale of variation (e.g., such as those suggested by the fibrils seen in H$\alpha$ high-resolution images) even a vertical magnetic field may modify significantly the linear polarization corresponding to the zero-field reference case. We find that the presence of a vertical field reduces the amplitude of the horizontal variation of the emergent $Q/I$ profile and produces $U/I$ signals. This occurs in both, the scattering geometry of a close to the limb observation and in the forward scattering geometry of a disk center observation. However, in this latter case a vertical field with strength in the Hanle saturation regime kills completely the horizontal fluctuation of the $Q/I$ signal, in addition to that of $U/I$.

(3) If the magnetic field in the above-mentioned scenario is significantly inclined (e.g., horizontal) and perpendicular to the orientation of the ``fibrils" we may have an increase in
the amplitude of the forward-scattering signals produced by the symmetry-breaking effects.


Finally, we point out that the results given in this paper in
graphical and tabular form, and the general symmetry and dimensional constrains derived in Sect.~\ref{sect24} and \ref{sect43}, serve as benchmarks for general
radiative transfer codes. With ours we plan to investigate how the presence
of inhomogeneities in the solar chromosphere may affect interpretations in terms of the Hanle effect
of the linear polarization signals in spectral lines like Ca {\sc ii} K, Mg {\sc ii} k and Ly{$\alpha$}.

\acknowledgements

Financial support by the Spanish Ministry of Science and Innovation through projects AYA2010-18029
(Solar Magnetism and Astrophysical Spectropolarimetry) and CONSOLIDER INGENIO CSD2009-00038 (Molecular Astrophysics: The Herschel and Alma Era) is gratefully acknowledged.

\appendix


\section{$\Lambda_{\alpha\beta}$ operators and their symmetry properties}
\label{app}

This appendix derives general expressions for the
$\Lambda_{\alpha\beta}$ operators introduced in equations
(\ref{eq53})-(\ref{eq55}) that give the explicit dependence 
of the radiation field
tensor components $J^K_Q$ 
on the statistical elements $S^K_Q$.
There is also a discussion on the symmetry properties of these operators according to the dimensionality of the problem.

The spherical components of the radiation field tensor 
can be written in a condensed manner as
\begin{equation}\label{jgen}
  J^K_Q = \int^\infty_0 \phi_\nu d\nu \oint \frac{d \Omega}{4 \pi}
  \sum_{i=0}^3 {\cal T}^K_Q (i, \Omega) I_i(\nu, \Omega),
\end{equation}
where $I_i(\nu, \Omega)$ ($i=0, 1, 2, 3$) stands for 
the Stokes parameters
$I_{\nu\vec{\Omega}}$, $Q_{\nu\vec{\Omega}}$, $U_{\nu\vec{\Omega}}$ 
and $V_{\nu\vec{\Omega}}$, respectively, while ${\cal T}^K_Q$ are
the polarization tensors introduced by Landi Degl'Innocenti (1984).
Analogously, the line source functions for each Stokes parameter
can be expressed as
\begin{equation}\label{slgen}
  S^{\rm line}_i \,=\,\sum_{KQ} w^{(K)}_{J_uJ_l} {\cal T}^K_Q (i, \vec{\Omega})
  S^K_Q,
\end{equation}
where $w^{(0)}_{J_uJ_l}=1$ and $w^{(2)}_{J_uJ_l}$ has been introduced after  
Eqs.~(\ref{eq02})
(atomic orientation is irrelevant in our problem and $w_{J_uJ_\ell}^{(1)}$ is not necessary here).
The total source function including the effect of an unpolarized background
continuum is then
\begin{equation}\label{sgen}
  S_i\,=\,r_\nu S^{\rm line}_i +(1-r_\nu) S^{\rm cont} \delta_{i, 0},
\end{equation}
where $\delta_{i, 0}$ is the Kronecker's delta 
and $r_\nu$ has been defined after Eqs.~(\ref{eq01}). 

From Eqs.~(\ref{eq01}), the Stokes parameters at a point $\mvec{x}$ in
the atmosphere can be expressed as follows (e.g., Mihalas 1978)
\begin{equation}
  [I_i(\nu, \mvec{\Omega})]_{\mvec{x}}=\,
	\int_{l(\mvec{x}, \mvec{\Omega})} S_i(\mvec{x}')
	{\rm e}^{-\tau(\mvec{x}, \mvec{x}')}
	\eta(\mvec{x}'){\rm d}s +
	I_i^{(0)}(x, \mvec{\Omega})
	{\rm e}^{-\tau(\mvec{x}, \mvec{x}_0)}
        =\Lambda_{\nu\mvec{\Omega}}
	[S_i]+{\cal I}_i(\nu, \mvec{\Omega}).
                                                          \label{bla}
\end{equation}
where the integral must be understood along the ray path
$\mvec{x}'=\mvec{x}'(s)$ with direction $\mvec{\Omega}$,
from the boundary point $\mvec{x}_0$ to $\mvec{x}$.
$I_i^{(0)}$ is the incident radiation field at $\mvec{x}_0$ and
$\tau(\mvec{x}, \mvec{x}')$ the optical depth between
$\mvec{x}$ and $\mvec{x}'$.
The second identity in Eq.~(\ref{bla}) introduces 
the integral operator $\Lambda_{x\mvec{\Omega}}$
and the attenuated Stokes parameters ${\cal I}_i$,

From Eqs.~(\ref{jgen})-(\ref{bla})
we get
\begin{multline}\label{a4}
  J^K_Q  = \,\int {\rm d}\nu\,\phi_\nu\,  \oint \frac{{\rm d} \mvec{\Omega}}{4
  \pi} \sum_{K' Q'} w^{(K')}_{J'J} \sum_{i=0}^3
  {\cal T}^K_Q (i, \mvec{\Omega}) {\cal T}^{K'}_{Q'} (i, \mvec{\Omega}) \Lambda_{\nu \mvec{\Omega}}
  [ r_\nu\,S^{K'}_{Q'} ] \\
   \quad +   \int {\rm d}\nu\,\phi_\nu\,  \oint \frac{{\rm d} \mvec{\Omega}}{4
  \pi} {\cal T}^K_Q (0, \mvec{\Omega}) \Lambda_{\nu \mvec{\Omega}}
  [ (1-r_\nu)\,S^{\rm cont} ] 
   \quad +  \int {\rm d}\nu \,\phi_\nu \oint \frac{{\rm d} \mvec{\Omega}}{4\pi} \sum_{i=0}^3
  {\cal 
  T}^K_Q(i, \mvec{\Omega}) {\cal I}_i(\nu, \mvec{\Omega}) ,
\end{multline}
where linearity of the $\Lambda_{\nu \vec{\Omega}}$ operator has been
applied.

Introducing the following $\Lambda$-like operators and 
attenuated radiation field tensor:
\begin{gather}\label{aa9}
  {\Lambda}_{KQ, K'Q'} [S^{K'}_{Q'}] =  \int {\rm d}\nu \,\phi_\nu\, \oint
  \frac{{\rm d} \mvec{\Omega}}{4 \pi}
  w^{(K')}_{J' J} 
  \sum_{i=0}^3 {\cal T}^K_Q (i, \mvec{\Omega}) {\cal T}^{K'}_{Q'} (i, \mvec{\Omega}) \Lambda_{\nu
  \mvec{\Omega}}[r_\nu\,S^{K'}_{Q'}] , \\
  {\Lambda}_{KQ, {\rm c}} [S^{\rm cont} ] =  \int {\rm d}\nu \,\phi_\nu\, \oint
  \frac{{\rm d} \mvec{\Omega}}{4 \pi}
  {\cal T}^K_Q (0, \mvec{\Omega}) \Lambda_{\nu \mvec{\Omega}}[(1-r_\nu)\,S^{\rm cont} ] , \\
{\cal J}^K_Q = \int \phi_\nu {\rm d}\nu \oint \frac{{\rm d} \mvec{\Omega}}{4\pi} \sum_{i=0}^3
  {\cal 
  T}^K_Q(i, \mvec{\Omega}) {\cal I}_i(\nu, \mvec{\Omega}),
\end{gather}
the radiation field tensor components can be expressed as
\begin{equation}\label{a9}
  J^K_Q = \sum_{K'=-2}^2\sum_{Q'=-K}^K {\Lambda}_{KQ,K'
  Q'}[S^{K'}_{Q'}] + {\Lambda}_{KQ,{\rm c}}[S^{\rm cont}]+{\cal J}^K_Q.
\end{equation}
Expressions such as these 
have been derived by
Landi Degl'Innocenti, Bommier \& Sahal-Br\'echot (1990).
However, it is more convenient from a numerical point of view to obtain similar 
expressions for the real (and linearly
independent) components introduced through Eqs.~(\ref{eq02_2})-(\ref{eq03_2}).
The expressions thus obtained ({Manso Sainz 2002}), can also be derived within 
the equivalent phase matrix formalism (e.g., Anusha \& Nagendra 2011).

\begin{table}
\begin{center}
  \caption{\label{tab4.3}}
  \begin{tabular}{l}
    \multicolumn{1}{c}{{\sc Analytical Expressions for} 
	$\varpi_{\alpha\beta}(\mvec{\Omega})$} \\
    \noalign{\smallskip}\hline\noalign{\smallskip}
    $\varpi_{00}=1$ \\
    $\varpi_{01}=\varpi_{10}= \frac{1}{2 \sqrt{2}}(3\mu^2 -1) $ \\
    $\varpi_{02}=2 \varpi_{20}= - \sqrt{3}\mu \sqrt{1-\mu^2} \cos{\chi}$ \\
    $\varpi_{03}=2 \varpi_{30}=  \sqrt{3}\mu \sqrt{1-\mu^2} \sin{\chi}$ \\
    $\varpi_{04}=2 \varpi_{40}=\frac{\sqrt{3}}{2} (1-\mu^2) \cos{2 \chi}$ \\
    $\varpi_{05}=2 \varpi_{50}=-\frac{\sqrt{3}}{2} (1-\mu^2) \sin{2 \chi}$ \\ 
	\noalign{\smallskip}\hline\noalign{\smallskip}
    $\varpi_{11}=\frac{1}{8} [(3\mu^2 -1)^2 +9 (1-\mu^2)^2]$ \\
    $\varpi_{12}=-2 \varpi_{21}=\sqrt{\frac{3}{8}}\mu \sqrt{1-\mu^2} \cos{\chi}
    [1-3\mu^2+3(1-\mu^2)]$ \\
    $\varpi_{13}=-2 \varpi_{31}=\sqrt{\frac{3}{8}}\mu \sqrt{1-\mu^2}\sin{\chi}
    [(3\mu^2-1)+3(\mu^2-1)]$ \\
    $\varpi_{14}=2 \varpi_{41}=\sqrt{\frac{3}{8}} (1+3\mu^2)(1-\mu^2) \cos{2 \chi}$ \\
    $\varpi_{15}=2 \varpi_{51}=-\sqrt{\frac{3}{8}} (1+3\mu^2)(1-\mu^2) \sin{2 \chi}$ \\
    \noalign{\smallskip}\hline\noalign{\smallskip}
    $\varpi_{22}=-\frac{3}{2}(1-\mu^2)(2 \mu^2 \cos^2{\chi}+\sin^2{\chi})$
    \\
    $\varpi_{23}=\varpi_{32}=\frac{3}{2}(2\mu^2-1)(1-\mu^2)\cos{\chi}\sin{\chi}$
    \\
    $\varpi_{24}=-\varpi_{42}=-\frac{3}{2} \mu \sqrt{1-\mu^2}(\cos{\chi}
    \cos{2\chi} \mu^2 +\sin{\chi}\sin{2\chi})$ \\ 
    $\varpi_{25}=-\varpi_{52}=\frac{3}{2} \mu \sqrt{1-\mu^2}(\cos{\chi}
    \sin{2\chi} \mu^2 -\sin{\chi}\cos{2\chi})$ \\ 
    \noalign{\smallskip}\hline\noalign{\smallskip}
    $\varpi_{33}=-\frac{3}{2}(1-\mu^2)(\cos^2\chi+2 \mu^2 \sin^2 \chi)$ \\
    $\varpi_{34}=-\varpi_{43}=\frac{3}{2}\mu\sqrt{1-\mu^2}(\mu^2 \cos 2\chi
    \sin\chi-\cos \chi \sin 2 \chi)$ \\
    $\varpi_{35}=-\varpi_{53}=-\frac{3}{2}\mu\sqrt{1-\mu^2}(\mu^2 \sin 2\chi
    \sin\chi+\cos \chi \cos 2 \chi)$ \\
    \noalign{\smallskip}\hline\noalign{\smallskip}
    $\varpi_{44}=\frac{3}{4}[\cos^2 2\chi (1+\mu^4)+2 \mu^2 \sin^2 2\chi]$
    \\
    $\varpi_{45}=\varpi_{54}=-\frac{3}{4} \cos 2\chi \sin 2\chi (\mu^2-1)^2$ \\
    \noalign{\smallskip}\hline\noalign{\smallskip}
    $\varpi_{55}=\frac{3}{4}[2\mu^2 \cos^2 2\chi + (1+\mu^4) \sin^2 2\chi]$
    \\
    \noalign{\smallskip}\hline
  \end{tabular}
\end{center}
\end{table}

To this end, we write 
the first term on the right hand side of Eq.~(\ref{a4})
\begin{multline}\label{extended}
  \int {\rm d} \nu \,\phi_\nu\, \oint \frac{{\rm d} {\vec{\Omega}}}{4
  \pi} \sum_{i=0}^3 {\cal T}^K_Q (i, \Omega)
  \Lambda_{\nu \Omega} \Big[ r_\nu\,\sum_{K'}
  w^{(K')}_{J' J} {\cal T}^{K'}_0 (i, \Omega) S^{K'}_0 +
   r_\nu\,\sum_{K'} \sum_{Q' >
  0} w^{(K')}_{J' J} \big[{\cal T}^{K'}_{Q'}(i, \Omega) S^{K'}_{Q'}+{\cal T}^{K'}_{-Q'}(i,
  \Omega) S^{K'}_{-Q'}\big]\Big].
\end{multline}
Since the ${\cal T}^K_Q(i, \vec{\Omega})$ tensor 
satisfies the conjugation 
property ${\cal T}^K_{-Q}=(-1)^Q[{\cal T}^K_Q]^*$, then ${\cal T}^{K'}_{-Q'}(i, \mvec{\Omega}) S^{K'}_{-Q'} = [{\cal T}^{K'}_{Q'}(i,\mvec{\Omega}) S^{K'}_{Q'} ]^*$.
Let $\tilde{\cal T}^K_Q$ and $\hat{\cal T}^K_Q$ 
be the real and imaginary parts of ${\cal T}^K_Q$, respectively.
The last bracket in
equation (\ref{extended}) then reads
\begin{equation}
  {\cal T}^{K'}_{Q'}(i, \mvec{\Omega}) S^{K'}_{Q'} + {\cal T}^{K'}_{-Q'}(i,
  \mvec{\Omega}) S^{K'}_{-Q'} =\, 2\,{\rm Re} \{{\cal T}^{K'}_{Q'}(i, \mvec{\Omega})
  S^{K'}_{Q'} \} 
	 = \,2\,\big[\tilde{\cal T}^{K'}_{Q'}(i, \mvec{\Omega}) \tilde{S}^{K'}_{Q'} -
  \hat{\cal T}^{K'}_{Q'}(i, \mvec{\Omega}) \hat{S}^{K'}_{Q'}\big].
\end{equation}
Finally, using this expression 
in Eq.~(\ref{extended}), and introducing the symbol $c_{Q'}=2-\delta_{Q'0}$,
we obtain the explicit dependence on $\tilde{S}^K_Q$ and $\hat{S}^K_Q$ of
the real and imaginary components $\tilde{J}{}^K_Q$ and
$\hat{J}{}^K_Q$ of the radiation field tensor:
\begin{align}
\begin{split}
  \tilde{J}^K_Q &= \int \,{\rm d} \nu \,\phi_\nu\, \oint \frac{{\rm d} \mvec{\Omega}}{4
  \pi} \sum_{K' Q'\geq 0} c_{Q'} w^{(K')}_{J' J} \Big\{\sum_{i=0}^3
  \tilde{{\cal T}}^K_Q(i) \tilde{\cal T}^{K'}_{Q'}(i) \Lambda_{\nu \mvec{\Omega}}[
  r_\nu\,\tilde{S}^{K'}_{Q'}] 
  -\sum_{i=0}^3 \tilde{\cal T}^K_Q(i) \hat{\cal T}^{K'}_{Q'}(i)
  \Lambda_{\nu \mvec{\Omega}}[r_\nu\,\hat{S}^{K'}_{Q'}] \Big\} \\&\quad
  + \int \,{\rm d} \nu \,\phi_\nu\, \oint \frac{{\rm d} \mvec{\Omega}}{4
  \pi} \tilde{\cal T}^K_Q(0) \Lambda_{\nu \mvec{\Omega}}[(1-r_\nu)\,S^{\rm cont}]
  + \tilde{{\cal J}^K_Q} ,
\end{split}
\displaybreak[0] \\
\begin{split}
  \hat{J}^K_Q &= \int {\rm d} \nu \,\phi_\nu \,\oint \frac{{\rm d} \mvec{\Omega}}{4
  \pi} \sum_{K' Q'\geq 0} c_{Q'} w^{(K')}_{J' J}
  \Big\{\sum_{i=0}^3 
  \hat{\cal T}^K_Q(i) \tilde{\cal T}^{K'}_{Q'}(i) \Lambda_{\nu \mvec{\Omega}}[
  r_\nu\,\tilde{S}^{K'}_{Q'}] 
	- \sum_{i=0}^3 \hat{\cal T}^K_Q(i) \hat{\cal T}^{K'}_{Q'}(i)
  \Lambda_{\nu \mvec{\Omega}}[r_\nu\,\hat{S}^{K'}_{Q'}] \Big\} \\ &\quad
  \int \,{\rm d} \nu \,\phi_\nu\, \oint \frac{{\rm d} \mvec{\Omega}}{4
  \pi} \hat{\cal T}^K_Q(0) \Lambda_{\nu \mvec{\Omega}}[(1-r_\nu)\,S^{\rm cont}]  
+ \hat{{\cal J}^K_Q},
\end{split}
\end{align}
which,  enlightening the notation, can be written as
\begin{align}
  \tilde{J}^K_Q & =  \sum_{K'Q'} {\Lambda}_{\widetilde{KQ},
  \widetilde{K'Q'}}[ \tilde{S}^{K'}_{Q'}] + \sum_{K'Q'}
  {\Lambda}_{\widetilde{KQ},
  \widehat{K'Q'}}[ \hat{S}^{K'}_{Q'}] +{\Lambda}_{\widetilde{KQ},
  {\rm c}}[ {S}^{\rm cont}]+\tilde{\cal J}^K_Q,
\displaybreak[0] \\
  \hat{J}^K_Q & = \sum_{K'Q'} {\Lambda}_{\widehat{KQ},
  \widetilde{K'Q'}}[ \tilde{S}^{K'}_{Q'}] + \sum_{K'Q'}
  {\Lambda}_{\widehat{KQ}, 
  \widehat{K'Q'}}[ \hat{S}^{K'}_{Q'}] +{\Lambda}_{\widehat{KQ},
  {\rm c}}[ {S}^{\rm cont}]+\hat{\cal J}^K_Q,
\end{align}
where 
\begin{align}
  {\Lambda}_{\widetilde{KQ},\widetilde{K'Q'}} &=  
  \int {\rm d} \nu \,\phi_\nu\, \oint \frac{{\rm d} \Omega}{4\pi} \,
  w^{(K')}_{J'J} \,c_{Q'} \,\varpi_{\widetilde{KQ},\widetilde{K'Q'}} \,
  \Lambda_{\nu \mvec{\Omega}}r_\nu\, ,   \label{eqa16}
\displaybreak[0] \\
  {\Lambda}_{\widetilde{KQ},c} &=  
  \int {\rm d} \nu \,\phi_\nu\, \oint \frac{{\rm d} \Omega}{4\pi} \,
  \varpi_{\widetilde{KQ}, {\rm c}} \,
  \Lambda_{\nu \mvec{\Omega}} (1-r_\nu)\,,  \label{eqa17}
\end{align}
and analogously for the remaining operators.
The angular weights are 
\begin{alignat*}{3}
  \varpi_{\widetilde{KQ},\widetilde{K'Q'}} &= 
  \sum_{i=0}^3 \tilde{\cal T}^{K}_{Q}(i) \tilde{\cal T}^{K'}_{Q'}(i), 
&\qquad\quad \varpi_{\widetilde{KQ},\widehat{K'Q'}} & = 
  -\sum_{i=0}^3 \tilde{\cal T}^{K}_{Q}(i) \hat{\cal T}^{K'}_{Q'}(i) ,
&\qquad\quad
\varpi_{\widetilde{KQ},{\rm c}} &= 
  \tilde{\cal T}^{K}_{Q}(0), \\
  \varpi_{\widehat{KQ},\widetilde{K'Q'}} & = 
  \sum_{i=0}^3 \hat{\cal T}^{K}_{Q}(i) \tilde{\cal T}^{K'}_{Q'}(i) ,   
&\qquad\quad
  \varpi_{\widehat{KQ},\widehat{K'Q'}} & = 
  -\sum_{i=0}^3 \hat{\cal T}^{K}_{Q}(i) \hat{\cal T}^{K'}_{Q'}(i) ,
&\qquad\quad
\varpi_{\widehat{KQ},{\rm c}} &= 
  \hat{\cal T}^{K}_{Q}(0).
\end{alignat*} 
Their explicit expressions are given in Tables~\ref{tab4.3} and \ref{tab4.4}
after a convenient index renaming (0, 1, 2, 3, 4 and 5 for
00, 20, $\widetilde{21}$, $\widehat{21}$, $\widetilde{22}$
and $\widehat{22}$, respectively).
\begin{table}
  \caption{\label{tab4.4}}
\begin{center} 
  \begin{tabular}{l}
    \multicolumn{1}{c}{{\sc Analytical Expressions}} \\
    \multicolumn{1}{c}{{\sc for}	$\varpi_{\alpha c}(\mvec{\Omega})$} \\
    \noalign{\smallskip}\hline\noalign{\smallskip}
    $\varpi_{0c}=1$ \\
    $\varpi_{1c}=\frac{1}{2\sqrt{2}} (3 \mu^2-1)$ \\
    $\varpi_{2c}=-\frac{\sqrt{3}}{2} \sin\theta \mu \cos\chi$ \\
    $\varpi_{3c}=-\frac{\sqrt{3}}{2} \sin\theta \mu \sin\chi$ \\
    $\varpi_{4c}=\frac{\sqrt{3}}{4} (1-\mu^2) \cos 2\chi$ \\
    $\varpi_{5c}=\frac{\sqrt{3}}{4} (1-\mu^2) \sin 2\chi$ \\
    \noalign{\smallskip}\hline
  \end{tabular}
\end{center}
\end{table}
Thus, we get (cf. Eqs.~(\ref{eq53})). 
\begin{subequations}\label{ap26}
\begin{align}
\mvec{\delta J^0_0}&=\mvec{\Lambda}_{00}\mvec{\delta S^0_0}+
\mvec{\Lambda}_{01}\mvec{\delta S^2_0}+
\mvec{\Lambda}_{02}\mvec{\delta \tilde{S}^2_1}+
\mvec{\Lambda}_{03}\mvec{\delta \hat{S}^2_1}+
\mvec{\Lambda}_{04}\mvec{\delta \tilde{S}^2_2}+
\mvec{\Lambda}_{05}\mvec{\delta \hat{S}^2_2}+
\mvec{\Lambda}_{0c}\mvec{\delta S^{cont}}+
\mvec{\delta {\cal J}^0_0}, 
\\
\mvec{\delta J^2_0}&=\mvec{\Lambda}_{10}\mvec{\delta S^0_0}+
\mvec{\Lambda}_{11}\mvec{\delta S^2_0}+
\mvec{\Lambda}_{12}\mvec{\delta \tilde{S}^2_1}+
\mvec{\Lambda}_{13}\mvec{\delta \hat{S}^2_1}+
\mvec{\Lambda}_{14}\mvec{\delta \tilde{S}^2_2}+
\mvec{\Lambda}_{15}\mvec{\delta \hat{S}^2_2}+
\mvec{\Lambda}_{1c}\mvec{\delta S^{cont}}+
\mvec{\delta {\cal J}^2_0}, 
\\
\mvec{\delta \tilde{J}^2_1}&=\mvec{\Lambda}_{20}\mvec{\delta S^0_0}+
\mvec{\Lambda}_{21}\mvec{\delta S^2_0}+
\mvec{\Lambda}_{22}\mvec{\delta \tilde{S}^2_1}+
\mvec{\Lambda}_{23}\mvec{\delta \hat{S}^2_1}+
\mvec{\Lambda}_{24}\mvec{\delta \tilde{S}^2_2}+
\mvec{\Lambda}_{25}\mvec{\delta \hat{S}^2_2}+
\mvec{\Lambda}_{2c}\mvec{\delta S^{cont}}+
\mvec{\delta \tilde{{\cal J}}^2_1}, 
\\
\mvec{\delta \hat{J}^2_1}&=\mvec{\Lambda}_{30}\mvec{\delta S^0_0}+
\mvec{\Lambda}_{31}\mvec{\delta S^2_0}+
\mvec{\Lambda}_{32}\mvec{\delta \tilde{S}^2_1}+
\mvec{\Lambda}_{33}\mvec{\delta \hat{S}^2_1}+
\mvec{\Lambda}_{34}\mvec{\delta \tilde{S}^2_2}+
\mvec{\Lambda}_{35}\mvec{\delta \hat{S}^2_2}+
\mvec{\Lambda}_{3c}\mvec{\delta S^{cont}}+
\mvec{\delta \hat{{\cal J}}^2_1}, 
\\
\mvec{\delta \tilde{J}^2_2}&=\mvec{\Lambda}_{40}\mvec{\delta S^0_0}+
\mvec{\Lambda}_{41}\mvec{\delta S^2_0}+
\mvec{\Lambda}_{42}\mvec{\delta \tilde{S}^2_1}+
\mvec{\Lambda}_{43}\mvec{\delta \hat{S}^2_1}+
\mvec{\Lambda}_{44}\mvec{\delta \tilde{S}^2_2}+
\mvec{\Lambda}_{45}\mvec{\delta \hat{S}^2_2}+
\mvec{\Lambda}_{4c}\mvec{\delta S^{cont}}+
\mvec{\delta \tilde{{\cal J}}^2_2}, 
\\
\mvec{\delta \hat{J}^2_2}&=\mvec{\Lambda}_{50}\mvec{\delta S^0_0}+
\mvec{\Lambda}_{51}\mvec{\delta S^2_0}+
\mvec{\Lambda}_{52}\mvec{\delta \tilde{S}^2_1}+
\mvec{\Lambda}_{53}\mvec{\delta \hat{S}^2_1}+
\mvec{\Lambda}_{54}\mvec{\delta \tilde{S}^2_2}+
\mvec{\Lambda}_{55}\mvec{\delta \hat{S}^2_2}+
\mvec{\Lambda}_{5c}\mvec{\delta S^{cont}}+
\mvec{\delta \hat{{\cal J}}^2_2},
\end{align}
\end{subequations}
where the matrices $\mvec{\Lambda}_{\alpha\beta}$ 
are obtained from $\mvec{\Lambda}_{\mvec{\Omega}}$ (see Eqs.~(\ref{eq54})-(\ref{eq55})).
%

\begin{figure}
  \centerline{\epsfig{figure=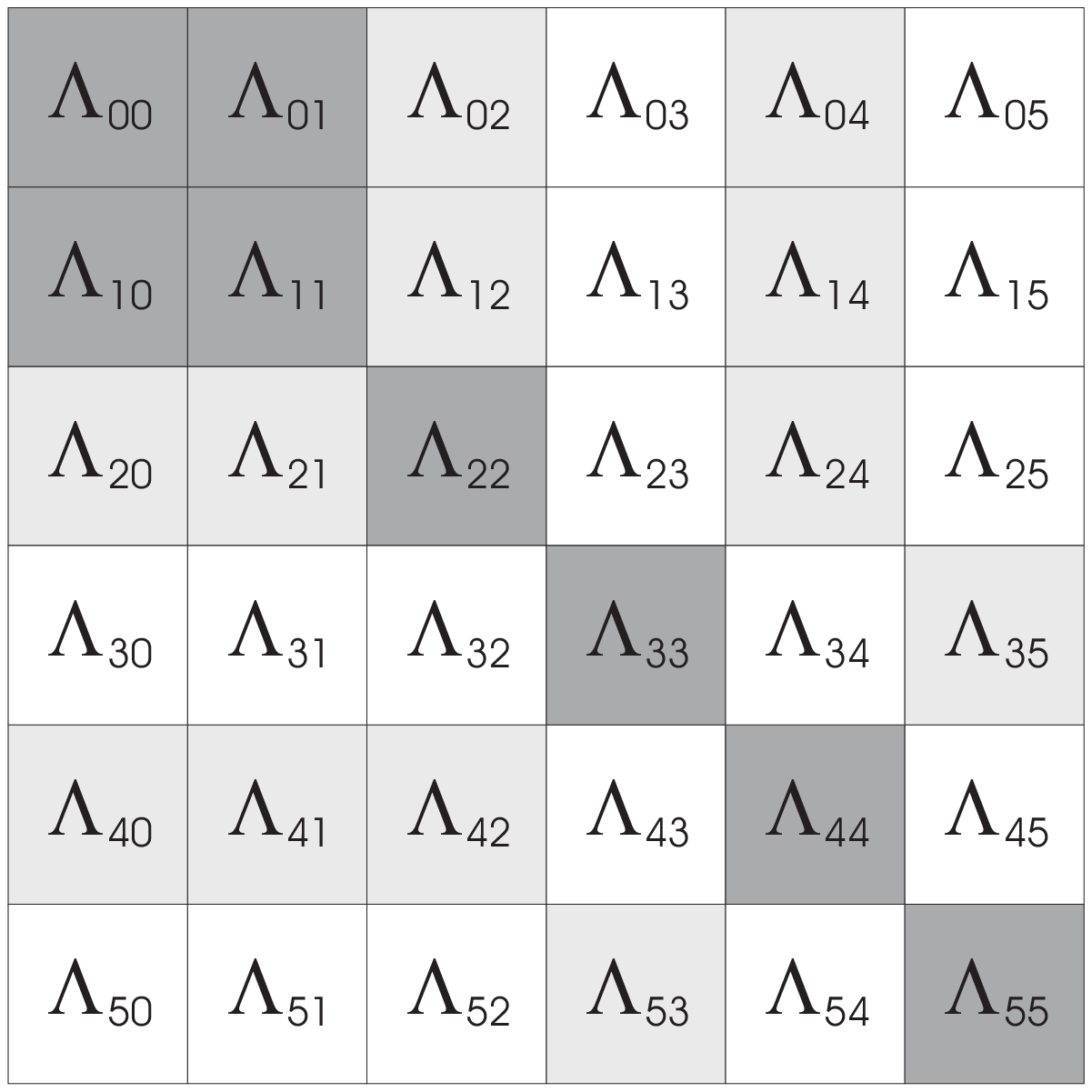, width=5cm}}
  \caption{\label{fig.4.7}
	Non-vanishing operators according to the medium's geometry.
	Dark blocks: plane-parallel case. Grey blocks: Cartesian
	two-dimensional medium. In a general three-dimensional atmosphere
	all of them are non-zero in general.}
\end{figure}

In a plane-parallel atmosphere, the ${\Lambda}_{\nu\mvec{\Omega}}(i, j)$
elements are independent of $\chi$ and 
the azimuthal integrals in Eqs.~(\ref{eqa16})-(\ref{eqa17}) can
be performed analytically. 
It is then found that all the non-diagonal
operators but $\mvec{\Lambda}_{01}$ and $\mvec{\Lambda}_{10}$, vanish,
as well as $\mvec{\Lambda}_{\alpha c}$ ($\alpha\ge 2$).
Furthermore, $\mvec{\Lambda}_{22}=\mvec{\Lambda}_{33}$ and $\mvec{\Lambda}_{44}=\mvec{\Lambda}_{55}$.

In a two-dimensional atmosphere, taking the $y$-axis as
the direction invariant under
translations, the $\mvec{\Lambda}_{\nu\mvec{\Omega}}$
operator satisfies the symmetry relation
\begin{equation}
   \Lambda_{\nu, (\mu, \chi)}(i, j)\,=\,\Lambda_{\nu, (\mu, -\chi)}(i, j).
                                                              \label{4.3d.7}
\end{equation}
Taking into account relation (\ref{4.3d.7}) in Eqs.~(\ref{eqa16})-(\ref{eqa17}), we find that 
$\mvec{\Lambda}_{03}$, $\mvec{\Lambda}_{05}$, $\mvec{\Lambda}_{13}$, 
$\mvec{\Lambda}_{15}$, $\mvec{\Lambda}_{23}$, $\mvec{\Lambda}_{25}$, 
$\mvec{\Lambda}_{34}$, $\mvec{\Lambda}_{45}$ and their symmetric ones
vanish, as well as $\mvec{\Lambda}_{3c}$ and $\mvec{\Lambda}_{5c}$.

These results are summarized in Fig.~\ref{fig.4.7}. 
Dark blocks show the non-zero operators in a plane-parallel atmosphere;
in a two-dimensional medium gray blocks are non-zero too.
Note that the $\hat{S}^2_1$ and $\hat{S}^2_2$ are radiatively coupled only 
between them.
Therefore, in the absence of magnetic field couplings, 
they are decoupled from the thermal source and vanish everywhere in the atmosphere.
This is another proof of the result obtained in \S 4.3 from
general symmetry considerations.
In a three-dimensional medium, all the operators are, in general,
non-zero.

\end{document}